\documentclass[article,5pt]{elsarticle}

\usepackage{color,soul}

\biboptions{authoryear}

\begin{document}

\begin{frontmatter}

\title{RadioLensfit: an HPC Tool for Accurate Galaxy Shape Measurement with SKA}

\author{Marzia~Rivi\corref{cor1}}
\ead{marzia.rivi@gmail.com}
\address{INAF - Istituto di Radioastronomia, via Gobetti 101, 40129, Bologna, Italy}

\author{Lance~Miller}
\address{Astrophysics, Department of Physics, University of Oxford, Keble Road, Oxford, UK}

\cortext[cor1]{Corresponding author}

\begin{abstract}
The new generation radio telescopes, such as the Square Kilometre Array (SKA), are expected to reach sufficient sensitivity and resolution to provide large number densities of resolved faint sources, and therefore to open weak gravitational lensing observations to the radio band.  
In this paper we present \textsc{RadioLensfit}, an open-source tool for an efficient and fast galaxy shape measurement for radio weak lensing shear.  It performs a single source model fitting in the Fourier domain, after isolating the source visibilities with a sky model and a \textit{faceting} technique. This approach makes real sized radio datasets accessible to an analysis in this domain, where data is not yet affected by the systematics introduced by the non-linear imaging process. We detail the implementation of the code and discuss limitations of the source extraction algorithm. We describe the hybrid parallelization MPI+OpenMP of the code, implemented to exploit multi-node HPC infrastructures for accelerating the computation and dealing with very large datasets that possibly cannot entirely be stored in the memory of a single processor. Finally, we present performance results both in terms of measurement accuracy and code scalability on SKA-MID simulated datasets. In particular, we compare shape measurements of 1000 sources at the expected source density in SKA Phase 1 with the ones obtained from the same dataset in a previous work by a joint fitting of the raw visibility data, and show that results are comparable while the computational time is highly reduced.
 
\end{abstract}


\end{frontmatter}


\section{Introduction}
\label{sec:intro}
Gravitational lensing is a phenomenon that causes the paths of photons arriving from distant sources to bend while they pass close to large matter accumulations because of the matter gravitational potential. The measure of this distortion on cosmological scales is a powerful technique for estimating the foreground mass distribution and the relationship between the distributions of dark and baryonic matter. Its combination with redshift measurements can provide cosmological constraints on the density of dark matter and, through the growth of large-scale structure, also on the dark energy component of the Universe. 
In the weak lensing regime, detectable around galaxy clusters at large angular distances, the images of galaxies behind clusters are weakly distorted (or sheared), resulting to be slightly elongated tangentially to the mass distribution. This anisotropic stretching is quantified in terms of gravitational shear that is directly accessible through measurements of galaxy shapes. Usually it is estimated considering only background star-forming (SF) galaxies under the assumption that they are randomly oriented. Then the shear is given by the weighted average of their ellipticities (see \cite{Kil15} for an overview). For a significant detection it is important to use a large sample of faint galaxies as the shear signal is very small (roughly a few percent at cosmological scales) compared with the intrinsic dispersion in galaxy shapes, typically around $\sigma_\varepsilon \sim 0.3$. For this reason weak lensing field has been served observationally only by deep optical surveys, owing to the larger number densities of faint galaxies achieved in such surveys and their higher resolution compared to other wavelengths.

An international effort is now ongoing to build the world largest radio telescope, called Square Kilometre Array\footnote{https://www.skatelescope.org} (SKA). The SKA will be developed over two main phases (SKA1 and SKA2), and will be partly hosted in South Africa (SKA-MID, mid-frequency dishes) and partly in Western Australia (SKA-LOW, low frequency stations). SKA-MID will have sufficient sensitivity to reach the needed SF galaxy number density for weak lensing measurements,  and a resolution to reliably measure galaxy shapes \citep{RedBook2018}. Although the faint sky observed for weak lensing surveys is SF galaxy dominated, in the radio band a non-negligible fraction of sources is expected to be associated with Active Galactic Nuclei (AGN)  \citep{COSMOS2017, GOODSN2018, RedBook2018}. This fraction is mainly formed by radio-quiet (RQ) AGN, which do not show jet-related emission and feature much weaker and compact radio emission \citep{Bonzini2013, Delvecchio2017, Mancuso2017}, therefore they could still be used for shear measurement. The remaining radio-loud (RL) AGNs is expected to be removed from the observed data due to their complicated morphologies, which would increase significantly the intrinsic shape dispersion (however an attempt to model the shape of RL AGN components and include them in the shear measurement has been made in \cite{Chang04}).
Weak lensing can thus become one of the primary science drivers in radio surveys too, with the advantage that they will access the largest scales in the Universe, probing to higher redshift than optical surveys such as LSST and Euclid \citep{Brown15}. Although the galaxy number density in the radio band will be lower than in the optical, the possibility to observe deeper can make radio weak lensing surveys for cosmology measurements competitive with the corresponding  optical surveys, as shown in recent forecasts from SKA simulations \citep{Harrison16}. Furthermore radio interferometry offers a deterministic knowledge of the instrument response, i.e. the Point Spread Function (PSF), and unique approaches that are not available at the optical band, such as concurrent measurements of polarization \citep{BB11} and galaxy rotation velocities \citep{HKEGS19}, that may be used to mitigate some of the astrophysical systematic effects encountered in weak lensing cosmology such as intrinsic alignments. Finally, cross-correlation of the shear estimators with optical surveys will also allow suppression of wavelength dependent systematic errors improving the measurement robustness and accuracy \citep{Patel10, DB16, Camera17}. 

The weak lensing community is currently planning a pilot continuum survey with SKA1-MID \citep{RedBook2018} to explore the actual potential of SKA-related weak lensing studies and to fine tune radio specific analysis methods in view of the full surveys that will be undertaken with SKA2.
Most of the techniques available for weak lensing shear measurement have been developed for optical surveys and are based on the model fitting of SF galaxies' shapes from images, e.g. \textsc{lensfit} \citep{Miller13} and \textsc{im3shape} \citep{im3shape}. In principle they can be applied to radio images as well, after correcting them to the radio galaxy model and the interferometer PSF. However radio observations measure \textit{visibilities}, i.e. the Fourier Transform (FT) of the sky image at a finite sampling (\textit{uv points}). They are usually transformed into images, corrected for the PSF, by an iterative non-linear process, which introduces residuals that may dominate the lensing signal \citep{Patel15}. A first attempt to apply an image-plane shape measurement method on the images obtained from radio data, involving a step to calibrate these biases, has been developed for the analysis of the data produced by the precursor radio weak lensing survey SuperCLASS \citep{SuperCLASS-III}. However, a more correct approach to adopt with radio data is to work directly in the visibility domain, where the noise is purely Gaussian and the data not yet affected by the systematics introduced by the imaging process. This approach is very challenging because sources are no longer localized in the Fourier domain and their flux is mixed together in a complicated way.  

In this domain, shape joint fitting of all sources in the field of view is obviously the most accurate way to take into account the signal contamination from nearby galaxies. Currently, two methods adopting this approach have been considered, assuming that source flux and positions are provided. The first one \citep{CR02} uses shapelets \citep{Refregier03, RB2003}, where galaxy images are decomposed through an orthonormal basis of functions corresponding to perturbations around a circular Gaussian. Shapelets are invariant under Fourier Transform (up to a rescaling) allowing the adoption in the visibility domain of the same approaches used with images. Finding the best fitting shape via minimising the chi-squared is nominally linear in the shapelet coefficients and therefore can be performed simultaneously for all sources by simple matrix multiplications. However, this is only the case once a size scale and number of coefficients to include have been chosen, which can in itself be a highly non-linear and time-consuming problem. Moreover shapelets introduce a shear bias as they cannot accurately model steep brightness profiles and highly elliptical galaxy shapes \citep{Melchior10}. For this reason, 2-component Sersic models are commonly used in optical weak lensing surveys analysis as they better describe SF galaxy surface brightness distribution \citep{Mandelbaum15}. In the radio band a reasonable assumption is to use a single optical disc-like component because the radio-emitting plasma should follow the distribution of stars in galaxy discs. This source model is used in the second approach \citep{rivi2019} where a joint fitting of both size and ellipticity parameters is performed by a Hamiltonian Monte Carlo (HMC) technique. HMC is a Markov Chain Monte Carlo method that adopts physical system dynamics to explore the likelihood distribution efficiently, resulting in faster convergence and maintaining a reasonable efficiency even for high dimensional problems \citep{Neal11}.  
Although this method provides very accurate results with simulated data, it is still very time-consuming even with a small sized dataset, and obviously convergence takes longer as the number of parameters increases. Therefore it could be unfeasible to deal with the expected real data size. 

In fact the computational challenges for a telescope such as SKA are great, given the large amount of visibilities that must be processed and the expected source number density. 
For example, a nominal weak lensing survey using the first 30\% of Band~2 will require at least 50~kHz frequency channel bandwidth, meaning about  6000 frequency channels, and  1 second integration time to make smearing effects tolerable \citep{SKA-ECP}. This means about 20~PB of raw visibilities per pointings of 1 hour integration time with the current design  of SKA-MID. This amount of data might be reduced by averaging visibilities in a grid but its effect on the galaxy ellipticity measurement is still to be investigated. Adding that the expected number of sources for such surveys is in the order of $10^4$ per square degree, it is clear that tools exploiting High Performance Computing (HPC) infrastructures are required for these data analysis.

In this paper we present \textsc{RadioLensfit},  a open-source tool which follows the approach adopted in optical surveys where each source is fitted once at a time after being extracted from the sky image. In particular it is an adaptation of \textsc{lensfit}, a method already used in optical weak lensing surveys such as CFHTLenS \citep{Heymans12}, KiDS \citep{Kuijken15, Giblin21}, RCSLenS \citep{Hildebrandt16a} and VOICE \citep{Fu18}, to the radio domain. \textsc{RadioLensfit} has been introduced with preliminary results on radio weak lensing shear measurement in \cite{Rivi16} and \cite{Rivi18}. Here, we describe the code implementation, with an optimized source extraction algorithm in the visibility domain (Section~\ref{sec:method}), and  its parallelization for the exploitation of multi-core and multi-node platforms (Section~\ref{sec:impl}). In Section~\ref{sec:results}, after tuning the source visibilities extraction algorithm, we show the method performance using the same SKA1-MID simulated dataset analyzed with HMC in \cite{rivi2019}, and in Section~\ref{sec:scalability} we present the code scalability over data size. Finally in Section~\ref{sec:real_data} we discuss the next steps towards the analysis of real data. Conclusions are summarized in Section~\ref{sec:end}.

\section{Software overview}
\label{sec:method} 

\textsc{RadioLensfit}\footnote{https://github.com/marziarivi/RadioLensfit2} is an open-source scalable C/C++ code for measuring star-forming galaxy shapes in the Fourier domain, where radio data originate. Following the approach used in optical surveys, for each source in the field of view the related visibilities are isolated given the source position and flux, and fitted independently. To further reduce nearby source signal contamination and accelerate computation, the model fitting is performed on a coarse uv grid after averaging data visibilities with sampling points falling in the same cells of the grid (see Section~\ref{sec:extraction}). As for \textsc{lensfit} \citep{Miller13}, the likelihood is marginalized over non-interesting source parameters and sampled on the ellipticity components using an adaptive grid around the maximum point. The main stages of the code are summarized below. 

\subsection{Data loading}
The standard data format for visibilities is the \textit{Measurement Set} (MS), which is organized in a table system. It was introduced with the widely used \textsc{CASA}\footnote{https://casa.nrao.edu/} software package for the post-processing (calibration, imaging and analysis) of radio astronomical data from interferometers, such as ALMA\footnote{https://almascience.eso.org/} and VLA\footnote{https://science.nrao.edu/facilities/vla}, as well as single dish telescopes.
We implemented a MS reader exploiting the \textsc{casacore}\footnote{https://casacore.github.io/casacore/} library.
A star-forming galaxy catalog must also be provided as a simple text file containing at least the position and the integrated flux at the reference frequency for each source. This catalog is expected to be ordered by decreasing signal-to-noise ratio (SNR), if provided, or flux, so that sources with higher SNR are the first to be processed.  

\subsection{Sky model computation}
Based on the information provided by the input source catalog, an approximation of the sky visibilities is computed and used as a first step for source extraction.  
Visibilities are evaluated at the interferometer baseline vectors, whose coordinates $(u,v,w)$ are measured in wavelengths at the center frequency of the channel bandwidth with respect to a coordinate system where $w$ axis points towards the phase tracking center~\citep{Thompson}. The sky model is defined as the sum of the visibilities of each source in the catalog. As mentioned in the introduction, SF galaxies in the radio band are modelled by an exponential brightness profile\footnote{The circular exponential surface brightness profile is $I(r) = I_0 \exp(-r/\alpha)$, where $I_0$ is the central brightness and $\alpha$ is the scalelength.  The exponential scalelength corresponds to a FWHM = $2\alpha \ln(2)$. The Gaussian profile is also provided, where FWHM = $2\sigma\sqrt{2\ln(2)}$.} (S\'ersic index 1) which should well describe the distribution of the synchrotron radiation emitted by the interstellar medium of the galaxy disc. The analytical Fourier Transform of this galaxy model has been computed in \cite{Rivi16}, here we also consider the $w$ coordinate which takes into account the effect of the non co-planar distribution of the antennae:
\begin{equation}
\label{model}
V(u,v,w) =  \Big( \frac{\lambda_\mathrm{ref}}{\lambda}\Big)^\beta  \frac{S_{\lambda_\mathrm{ref}}\mathrm{e}^{2\pi \mathrm{i}\bigl(u l + v m+w (\sqrt{1-l^2-m^2}-1)\bigr)}}{\big(1+4\pi^2 \alpha^2 |\mathbf{A}^{-T}\mathbf{k}|^2\big)^{3/2}}.
\end{equation}
$\beta$ is the spectral index for which we adopt a fiducial value of $-0.7$, $\mathbf{k}$ is the vector $(u,v)$, and the remaining are source parameters: image position coordinates $(l,m)$, integrated flux density~$S_{\lambda_\textrm{ref}}$ at reference wavelength $\lambda_\textrm{ref}$, scalelength~$\alpha$ and ellipticity components $e_1, e_2$. The shape distortion according to the ellipticity parameters is introduced by the matrix~$\mathbf{A}$ that linearly transforms the circular exponential profile to an ellipse:
\begin{equation}\label{linearTransf}
\mathbf{A}  = \left( \begin{array}{cc} 1-e_1 & -e_2 \\ -e_2 & 1+e_1 \end{array} \right).
\end{equation}
We assume the following ellipticity definition:
\begin{equation}
\label{e1e2}
\mathbf{e} = e_1 +\mathrm{i}e_2 = \frac{a-b}{a+b}\mathrm{e}^{2\mathrm{i}\theta},
\end{equation}
where $a$ and $b$ are the galaxy major and minor axes respectively, and $\theta$ is the galaxy orientation (position angle).   
The initial sky model is defined assuming that all sources are circular (meaning $e_1=e_2=0$), but is refined each time a source is fitted, by replacing the ellipticity of the source model with the measured one.
The galaxy scalelength can either be provided in the source catalog or estimated from $S_{\lambda_\textrm{ref}}$ according to the following linear relation between the log of the median value $\alpha_\mathrm{med}$ and the flux density $S$:
\begin{equation}
\ln{[\alpha_\mathrm{med}/\textrm{arcsec}]} = -0.93 +0.33\ln{[S/\mu \textrm{Jy}]}.
\label{scale-flux}
\end{equation}
This relation has been estimated in \cite{Rivi15} by the analysis of the publicly available source catalog\footnote{https://heasarc.gsfc.nasa.gov/W3Browse/radio-catalog/vlasdf20cm.html} of the VLA SWIRE Deep Field 20 cm survey. 
The coefficients of this relation are respectively the algorithm parameters \texttt{ADD} and \texttt{ESP}. They can be tuned by the analysis of more recent source observations with the new generation of radio telescopes.

\subsection{Source extraction}
\label{sec:extraction}
The sky model visibilities are used to remove from the observed data the approximated visibilities of all the sources in the field of view except the one we want to isolate for shape fitting.  
For this reason, ordering the source extraction by decreasing SNR/flux allows to isolate sources at low SNR with a better approximation of the sky model. The extraction is completed using a \textit{faceting} technique that further reduces the galaxy signal contamination by what remains of the other sources signal: given the position of the source to be extracted, galaxy visibilities are first phase shifted, by multiplying each visibility by  the term $\exp(-2\pi \mathrm{i} (u l + v m+w (\sqrt{1-l^2-m^2}-1)\bigr))$, in order to move the pointing center to the location of the source in the image domain, and then averaged in a coarse square regular uv grid (\textit{facet}). This way we reduce the field of view to a small patch around the source and reduce the number of visibilities used for the model fitting, accelerating the computation. The $w$ coordinate can now be neglected due to the small field of view of the facet.

With respect to \cite{Rivi18}, where the size of the facet is constant for all the sources with flux in a fixed range, here  
it is defined per source as follows. If the source scalelength at the reference frequency is not known, it is estimated from the source flux according to equation~(\ref{scale-flux}) as for the sky model, then it is multiplied by a constant factor $K$ to define a field of view sufficiently large to contain entirely the source (intrinsically convolved with the PSF).
The number of cells of the facet is obtained by applying the relation between the field of view $\theta$ (in radians)  and the related uv grid cell size (in units of wavelengths) $\Delta_u = 1/\theta$ \citep{Weighting99}. The $K$ factor is another algorithm parameter (called \texttt{PSF\_NAT}) dependent on the distribution of the uv points (uv coverage). 
As discussed in \cite{Rivi18}, in the fitting phase we do not model exactly the synthesized beam of the source model visibilites because they are evaluated directly at the uv points of the facet. Therefore the facet field of view should be a good trade-off between to be large enough to contains all the significant source signal but sufficiently small to suppress source model side-lobes which are not suppressed by any apodization as for the averaged data (see Section~\ref{sec:facet}). In this respect, defining a facet with source-dependent size allows to adapt it as much as possible to each source.  
Moreover it may also reduce significantly the cost of the model fitting, compensating the computational cost of the facet size and uv coordinates re-computation. 

\subsection{Source fitting} 
\label{uvfit}
The facet model visibilities are computed at the facet uv points as in equation~(\ref{model}) but with $w=0$, and we adopt a chi-squared approach for the computation of the likelihood.
The likelihood is marginalized over non-interesting parameters such as position, flux and scalelength in order to obtain a function of only the ellipticity components. Marginalization over flux and position is computed analytically and semi-analytically respectively, adopting a uniform prior as described in \cite{Rivi16}. Marginalization over scalelength is performed numerically over a finite interval [\texttt{RMIN}, \texttt{RMAX}] with \texttt{NUM\_R} quadratic spacings, assuming a log-normal prior \citep{Rivi15} with flux-dependent mean given by equation~(\ref{scale-flux}) and default standard deviation \texttt{R\_STD}~=~0.3136.  The numerical marginalization is performed using the interpolation and quadrature functions of the GNU Scientific Library (\textsc{GSL}).
The likelihood maximum point is computed using the simplex method \citep{Nelder-Mead} with tolerance \texttt{TOL}~=~$10^{-3}$ and starting point (0,0). This maximization method does not require the knowledge of the likelihood gradient and is provided by the \textsc{GSL} library.
The galaxy ellipticity measurement is obtained after sampling the likelihood with an adaptive grid covering a neighbourhood around the maximum point. The sampling terminates when either a maximum number \texttt{NP\_MAX}~=~30 of samples has been measured above a threshold of 5\% of the maximum likelihood, or a resolution \texttt{E\_RES}~=~0.003 in ellipticity component is reached. As measure of the galaxy's ellipticity components we take the mean and standard deviation of the ellipticity samples above the threshold.

\begin{figure}
\centering
\includegraphics[scale=0.48]{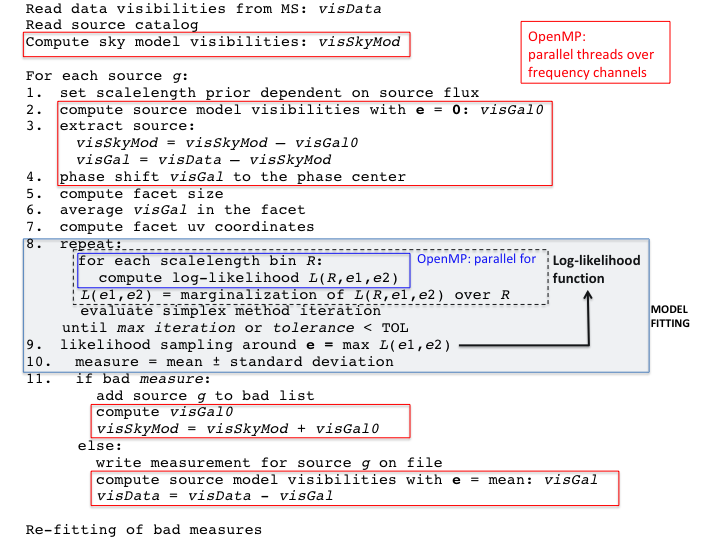}
\caption{Sketch of the serial code (+ OpenMP).}
\label{fig:sketch-code}
\end{figure}

Unreliable fits are recognised by a too small variance of the ellipticity likelihood to be realistic. This may be related to a likelihood not sufficiently smooth, leading to local maxima of the sampled region or errors in the cross-correlation marginalizaton over source position and scalelength, possibly due to PSF sidelobes or too much noise in the data. The variance threshold \texttt{VAR} for selecting these ”bad measurements” is another algorithm parameter that can be defined based on the maximum source SNR. 
The corresponding sources are not removed in the first instance from the data and sky model visibilities, but they are re-fitted at the end of the procedure, when the ellipticities of all the other sources are measured and a better sky model is obtained.  

\medskip
All the above algorithm parameters are set and can be modified in the file \texttt{default\_params.h}. 
The code returns a text file containing a new source catalog with the ellipticity components measurements added.
A sketch of the algorithm is shown in Fig.~\ref{fig:sketch-code}, where OpenMP parallel sections for accelerating the computation are highlighted (this level of parallelization is discussed in Section~\ref{sec:OpenMP}).

As discussed in the introduction, to manage the overwhelming volume of real SKA raw (calibrated) visibilities, we need a further level of code parallelization to distribute the data among different processors to perform source extraction in a reasonable time (see Section~\ref{sec:scalability}). For this reason we have implemented a hybrid parallelization MPI+OpenMP of the code as described in the next section.
 
 \section{Parallelization}
\label{sec:impl}
The code has two levels of parallelization that can be combined together in a hierarchical way: 
a) \textit{multi-node}, using the library Message Passing Interface (MPI\footnote{https://www.mpi-forum.org}); it allows to deal with very large datasets, especially for those that cannot be processed in the memory of a single node of the platform, and at the same time to perform the model fitting of more sources simultaneously (see Section~\ref{sec:MPI});
b) \textit{multi-core}, using the API OpenMP\footnote{https://www.openmp.org/} and exploiting the shared memory of the processor; it accelerates the most computationally intensive and embarrassingly\footnote{Meaning that there is no dependency or need for communication among the parallel sections.} parallel parts of the code, such as the computation of the model visibilities, as described in Section~\ref{sec:OpenMP}.

\subsection{MPI}
\label{sec:MPI}

\begin{figure}
\centering
\includegraphics[scale=0.76]{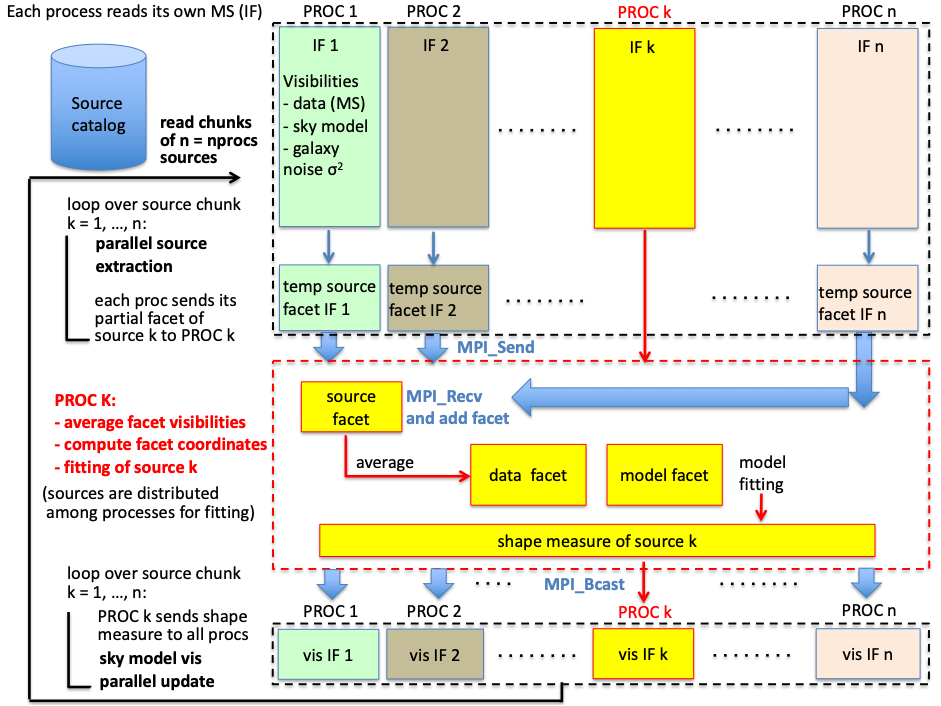}
\caption{Sketch of the MPI parallelization.}
\label{fig:sketch-mpi}
\end{figure}

The higher level of parallelization of the code consists in splitting large MS in sub-MS each containing an Intermediate Frequency (IF) band (i.e. spectral window) to be processed by a different MPI task (process) on a separate node of the system. Each process reads its own MS and stores in the node memory the visibilities of the original data, sky model and single source for its own fraction of frequencies. 
Source extraction is performed jointly by all tasks each working on its own data and storing the facet visibilities of the current source for its IF band in a temporary array.
Extracted sources are distributed among MPI processes so that each task performs simultaneously the model fitting of a different source. For this reason source extraction occurs in chunks of length equal to the number of MPI tasks.
The contribution to the facet visibilities related to the other spectral windows are collected from the other MPI tasks by MPI point-to-point\footnote{MPI point-to-point functions involve only two processes. We do not use collective communication functions in this step in order to save memory.} communication functions and then averaged. 
At the end of the source fitting, the task sends the shape measurement to all other processes as they have to perform jointly the update of the sky model using the measured ellipticity as new shape parameter of the source. Fig.~\ref{fig:sketch-mpi} shows a sketch of this parallelization.  
This way we have minimized communication among processes as well as the overhead due to the message passing latency. Workload may not be well balanced among processes as the source fitting time may significantly vary depending on source. However, as sources are extracted sequentially according to SNR (or source flux) order, the sources in the same chunk should take similar computing time for model fitting. 

Note that due to the parallel fitting of the sources, the sky model now can be updated only in chunks of sources and therefore there could be a little discrepancy with the results of the serial version of the code.  

\subsection{OpenMP}
\label{sec:OpenMP}
In the second level of parallelization, each MPI task forks in OpenMP threads in the parallel regions, exploiting all the cores of the node for most of the computation. 
This parallelization consists in using OpenMP C directives defining sections of the code where parallel threads are created at the beginning of each section and destroyed at the end. The number of threads are usually defined as the number of cores within the node so that a single thread works on each core, but 2 or 4 threads can also work on the same core if hyper-threading is enabled. Threads share the node memory so they can directly access the same arrays but work on different elements. We use the \texttt{parallel for} directive to parallelize the computation of the most external loops of the code with independent iterations, as highlighted in Fig.~\ref{fig:sketch-code}.
In the sky model computation and source extraction phase, the visibilities computation is distributed among threads each working for different frequencies of the spectral window\footnote{Averaging of the visibilities is not parallelized because there could be a concurrent access to the facet array elements, being facet coordinates measured in wavelengths units.}. 
In the model fitting phase each thread computes the likelihood (including source model visibilities and marginalization over source position) for different scalelength values of the source model. These likelihood values are then used for the numerical likelihood marginalization over the scalelength parameter.
 
\section{Method performance} 
\label{sec:results}

In this section we investigate the performance of the method on SKA1-MID simulated data in terms of shape bias and computing time.
We assume the linear bias model for comparing input and measured ellipticity values:
$$\tilde e_i - e_i = m_i e_i + c_i, \qquad i=1,2,$$
where $\tilde e_i$ (resp. $e_i$) is the $i$-component of the measured (resp. original) value of the input ellipticity, as defined in equation~(\ref{e1e2}), $m_i$ and $c_i$ are respectively the multiplicative and additive biases.
As for optical observations, multiplicative bias arises from  effects such as noise bias and neighbour bias (due to source signal contamination by residuals of nearby galaxies). If we consider the weak lensing shear obtained as a weighted average of sources' ellipticities, other effects such as source selection and weight are contributing to the multiplicative bias. Additive bias indicates a systematic error due to effects such as the correlation of noise bias with the PSF anisotropy that coherently smears the source shape. For an overview on these effects and their contribution on the bias on optical surveys see \cite{Miller13,mandelbaum18,Samuroff18,Zuntz18}. In the radio band we expect a similar behaviour, and similar approaches for their mitigation may be implemented quite easily in the interferometer data analysis, with the advantage that for the additive bias at radio wavelengths we know the PSF much better than at optical wavelengths and we can make the PSF isotropic directly by weighting the uv-plane and with a suitable choice of the integration time. Moreover, in the Fourier domain the neighbour bias may be almost removed by a joint fitting of all sources in the field of view such as with the HMC approach adopted in \cite{rivi2019}. We then use the same simulation ("SKA1 1000" dataset) and the optimal results obtained in such paper as a reference for evaluating the performance of \textsc{RadioLensfit}, where the faceting in the visibility domain for source extraction and model fitting may significantly affect both multiplicative and additive biases (see Section~\ref{sec:facet} and \cite{Rivi18}).
 
The "SKA1 1000" dataset was simulated using the SKA-MID Phase 1 antennae configuration, and assuming no time and frequency smearing effects so that a reduced uv coverage\footnote{See Fig.~2 of \cite{rivi2019}.} of a real 8-h observation pointing close to the zenith was generated considering only a single large frequency channel (240 MHz of bandwidth) centered at 1.4~GHz and an integration time $\Delta t = 60$~s. Observed visibilities were directly simulated\footnote{Note that visibilities sampled at the points of the uv coverage are the Fourier Transform of the observed dirty image, where the sources are convolved with the radio telescope PSF.} from the source catalog, using equation~(\ref{model}) and adding an uncorrelated\footnote{See appendix in \cite{Rivi16}.} Gaussian noise, whose variance is dependent on the antenna system equivalent flux density (SEFD) of SKA1 dishes, the frequency channel bandwidth and the time sampling. We refer to Section~3 in \cite{Rivi18} for more details, as well as for a description of the generation of source catalogs according to realistic parameters distributions. 

\subsection{Facet size tuning}
\label{sec:facet}
Facet field of view factor $K$, as defined in Section~\ref{sec:extraction}, is dependent on the uv coverage of the observation, i.e. the PSF with which the source is convolved in the image domain. Therefore, before using the tool is recommended to tune the optimal facet size by testing it on simulations performed with the same uv coverage of the original dataset.

We simulated a realistic population of 1000 faint SF galaxies, a single source in the field of view at a time, with flux ranging between 200~$\mu$Jy and 10~$\mu$Jy, and a reduced instrumental noise in order to have a sufficient SNR to emphasise only facet size effects in the shape measurement. The realistic noise variance was reduced by a factor~5, corresponding to a minimum source SNR~$\sim 19$. 

\begin{figure}
\centering
\includegraphics[scale=0.5]{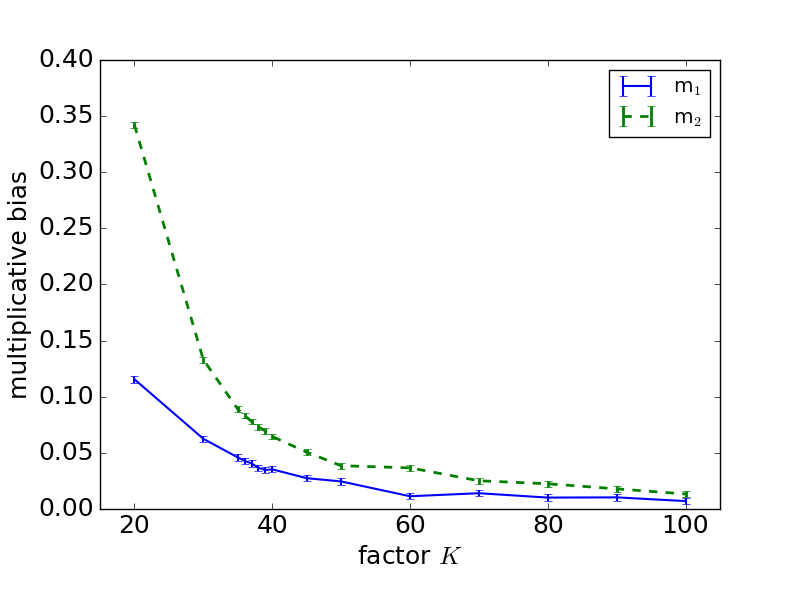}
\includegraphics[scale=0.5]{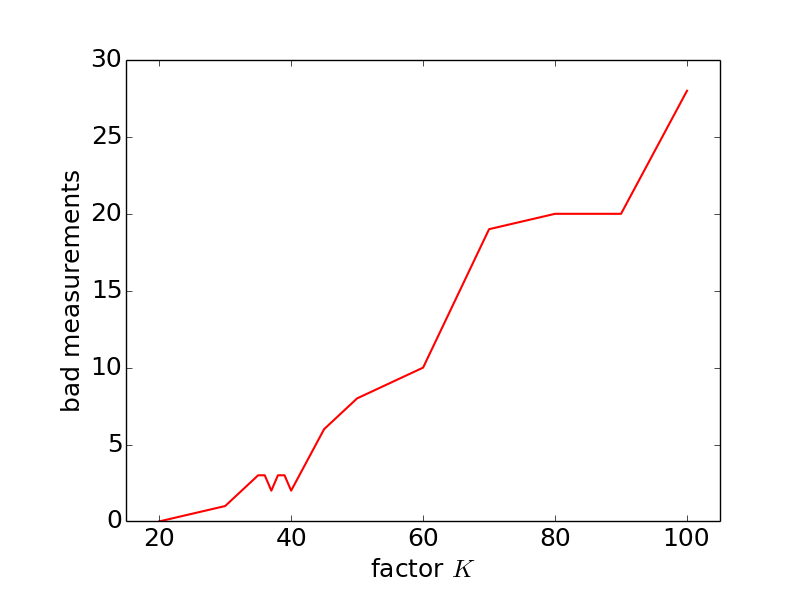}
\caption{Plots of the multiplicative bias (\textit{upper panel}) and the number of bad estimates (\textit{lower panel}) as functions of the facet size factor $K$, for the shape fitting of a realistic distribution of 1000 single sources in the field-of-view.}
\label{fig:slopes}
\end{figure}

Fig.~\ref{fig:slopes} shows the multiplicative bias estimates obtained from the best-fitting lines of source ellipticity measurements for different $K$ values. Additive biases are of the order of $10^{-3}$. 
Note that for all the facet sizes the measurement of the first ellipticity component is always better than the second component, this is probably due to the slight anisotropy of the instrumental PSF (i.e. asymmetry of the uv coverage) combined with the facet size.

\begin{figure}
\centering
\includegraphics[scale=0.285]{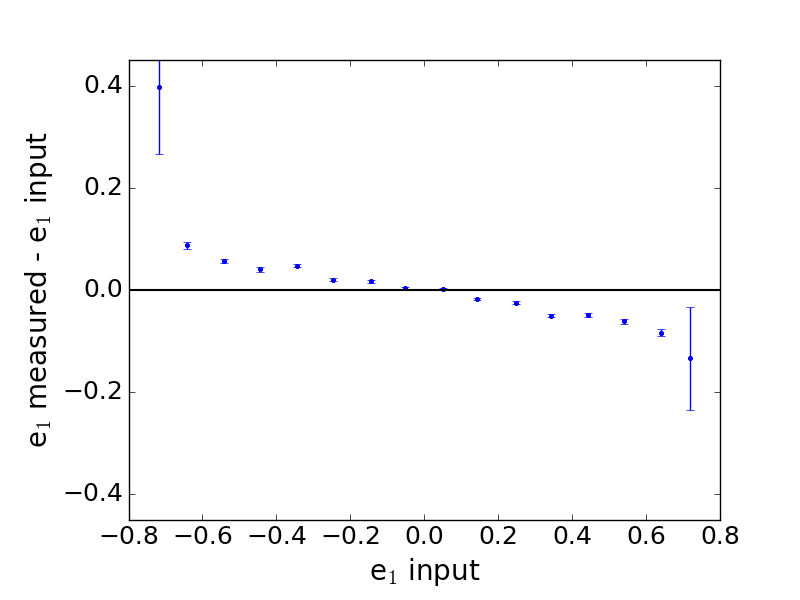}
\includegraphics[scale=0.285]{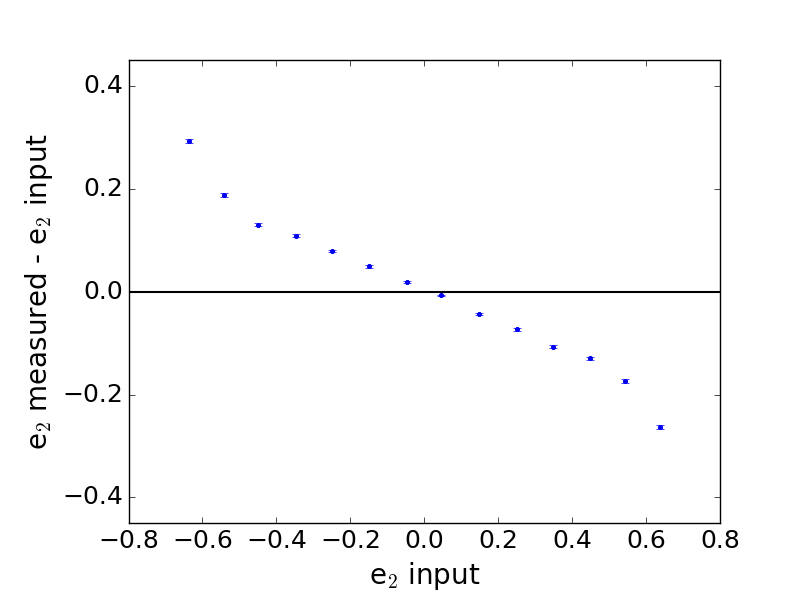}
\caption{Binned ''measured minus true'' values, $\Delta e_i$, of  both ellipticity components of galaxies with SNR $> 19$ for $K=20$. As emphasised here, facets too small do not contain entirely the signal of highly elliptical sources (convolved with the PSF) producing a distorted shape fitting.}
\label{fig:K20}
\end{figure}

\begin{figure}
\centering
\includegraphics[scale=0.285]{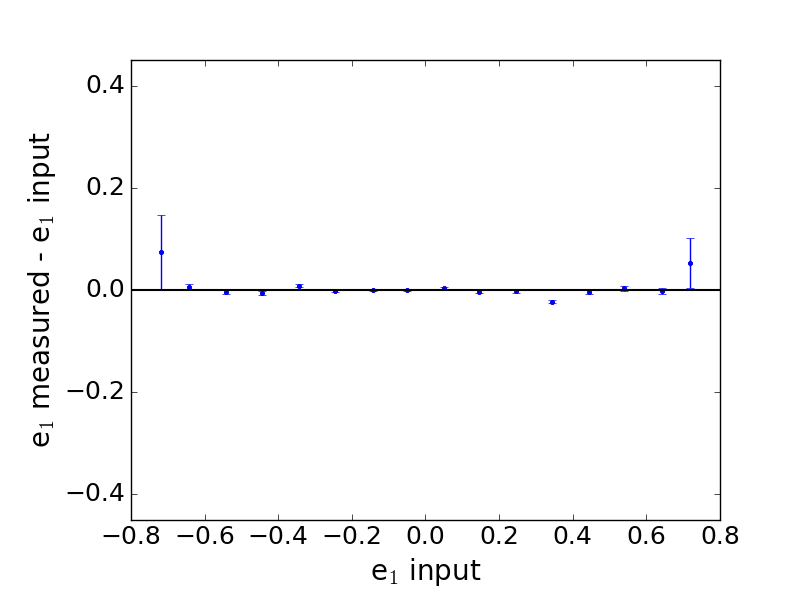}
\includegraphics[scale=0.285]{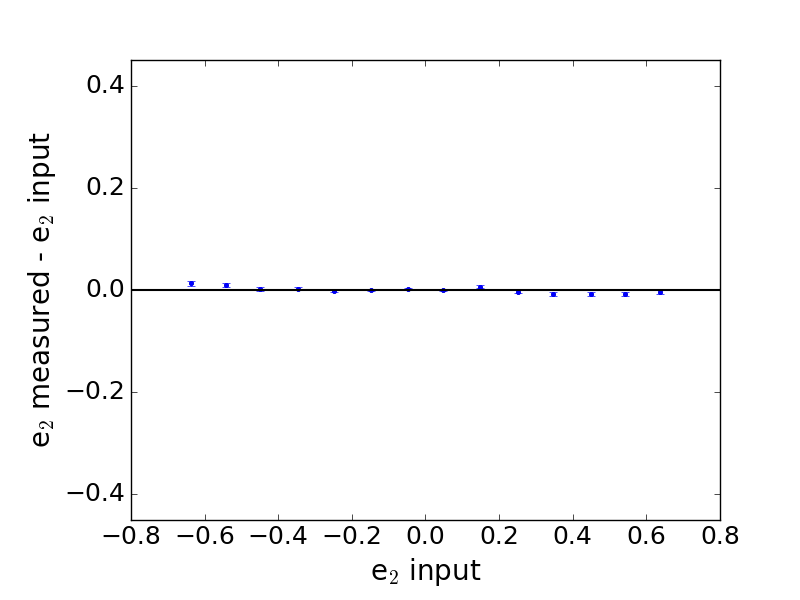}
\includegraphics[scale=0.285]{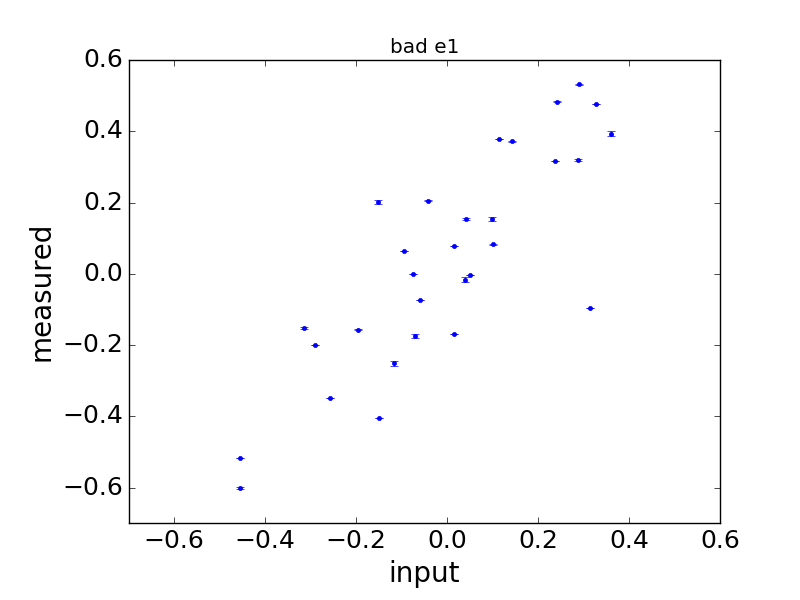}
\includegraphics[scale=0.285]{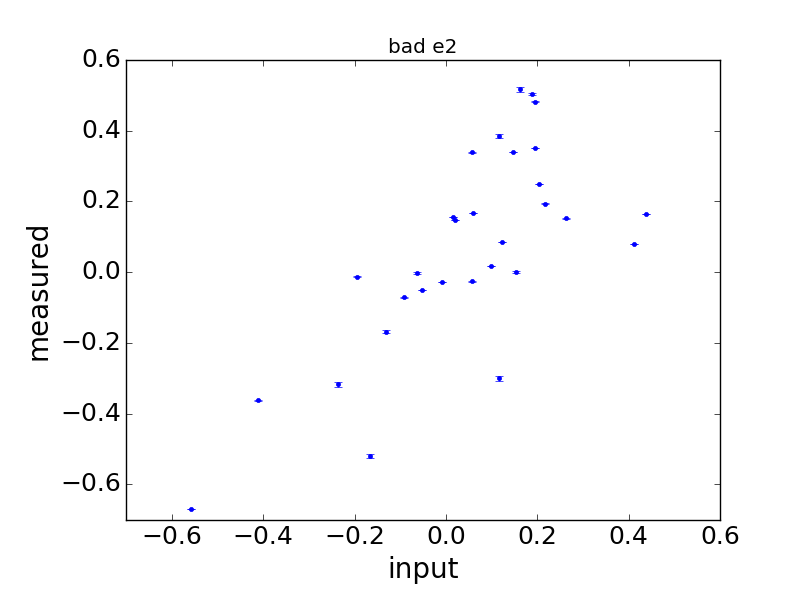}
\caption{Binned ''measured minus true'' values, $\Delta e_i$, of  both ellipticity components of galaxies with SNR $> 19$ for $K=100$. Although the facet is sufficiently large to contain the sources entirely (\textit{upper panel}), the discrepancy between the synthesized beam of the data and source model visibilities produces 28 bad measurements independently of the source shape (\textit{lower panel}).}
\label{fig:K100}
\end{figure}

As expected, for too small facets we observe that on average the difference between the measured and true values is larger for highly elliptical sources. See Fig.~\ref{fig:K20} for the extreme case of $K=20$. This is not due to the difficulty of the model fitting (which is well known to be higher for round sources, as shown in~\cite{rivi2019}), but by the fact that elliptical large sources are not fully contained in the facet. 
On the other side, we observe that the minimum size required to contain all the source signal for shape measurement may be too large to mitigate the synthesized beam model discrepancy\footnote{The averaging of data visibilities falling in the same facet cells almost remove side-lobes effects which are different and still present in the model visibilities, being directly evaluated at the center point of each facet cell.} at low SNR. In fact the likelihood may be so noisy in the fitting of some sources that either the likelihood maximization or sampling may return unacceptable values as measurement of the shape.
We remove these bad results, after re-fitting, when the standard deviation $\sigma$ of an ellipticity component or the 1D variance\footnote{Square root of the covariance matrix determinant.} $\bar \sigma^2$ of the measure is still too strict to be acceptable (e.g. for a dynamical range of the SNR between 400 and 8 we require at least $\sigma(e_i) > 5 \cdot 10^{-3}$ and $\bar \sigma^2(\mathbf{e}) > 5 \cdot 10^{-5}$). For example, for $K=100$ we observe that $2.8\%$ of the measurements are bad independently of their ellipticity (see lower panel of Fig.~\ref{fig:K100}). The fitting of the remaining sources is good, as showed in the upper panel of Fig.~\ref{fig:K100}, where measurements are on average close to the true ones. 

\begin{figure}
\centering
\includegraphics[scale=0.285]{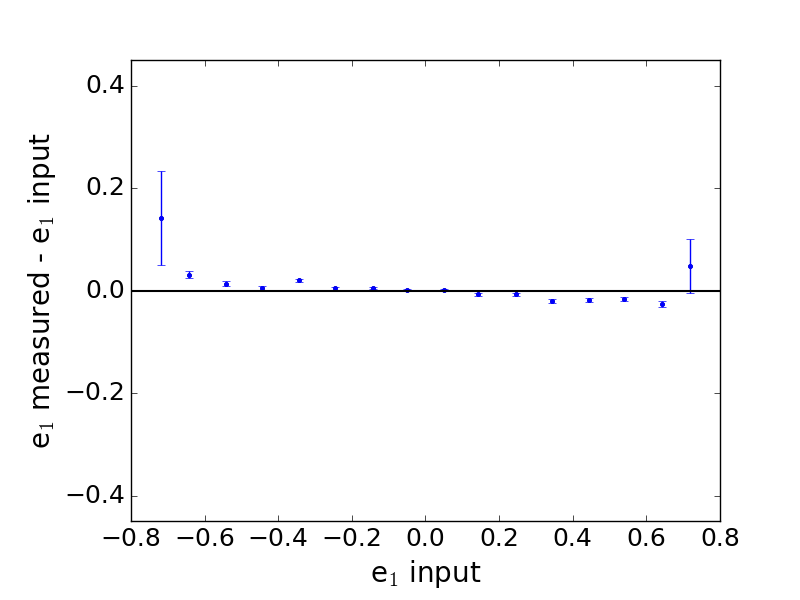}
\includegraphics[scale=0.285]{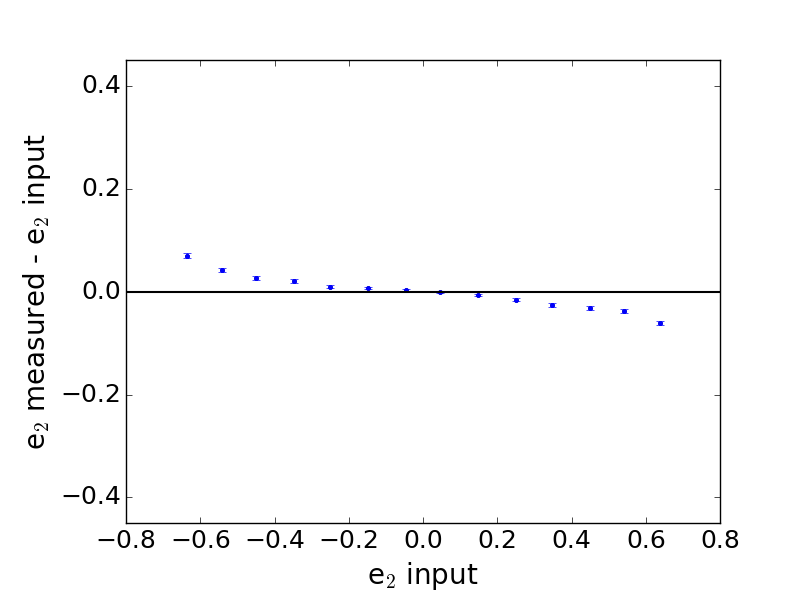}
\caption{Binned ''measured minus true'' values, $\Delta e_i$, of  both ellipticity components of galaxies with SNR $> 19$ for $K=38$. On average the discrepancy is larger than 0.05 only for $|e| > 0.6$.}
\label{fig:K38}
\end{figure}

We choose $K = 38$ as a good trade-off between a reduced statistics (due to the loss of shape measurements) and shape accuracy, as well as to minimize the facet size in order to remove most of nearby source signal contamination in real observations. Fig.~\ref{fig:slopes} shows that the multiplicative bias of both ellipticity components  starts to flatten from this value, for which we obtain $ m_1 = 0.0366 \pm 0.0028$ and $m_2 = 0.0731 \pm 0.0026$, and the number of bad measurements is only 0.3\%. Moreover, on average the discrepancy due to a possible shape ''cutting'' is larger than 0.05 only for highly elliptical sources, i.e. for $|e| > 0.6$ (see Fig.~\ref{fig:K38}) corresponding to a negligible number of galaxies according to the probability distribution of the ellipticity module presented in~\citep{Miller13}. This choice also reduces the computational cost considerably.

\subsection{Shape bias}
\label{ska1-mid-sim}

\begin{figure}
\centering
\includegraphics[scale=1.3]{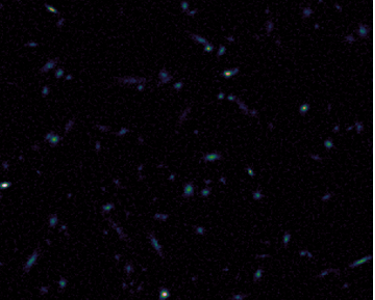}
\caption{Zoom in the clean image of the simulated sky visibilities.}
\label{fig:image}
\end{figure}

We now use the simulated dataset  "SKA1 1000"  described in~\cite{rivi2019} to compare  \textsc{RadioLensfit} results with the ones produced by the HMC joint fitting of all the 1000 sources in the field of view at the density expected for SKA-MID continuum surveys at 1.4~GHz in Phase~1, i.e. 2.7~gal/armin$^2$. Fig.~\ref{fig:image} shows a fraction of the clean image of the simulated sky visibilities. The source population, whose flux is still ranging between 200~$\mu$Jy and 10~$\mu$Jy, has a SNR $\ge 8.5$ (of which 26 galaxies have SNR below 10) because a realistic instrumental noise is added. Source scalelength parameter is ranging between \texttt{RMIN} = 0.3 arcsec and \texttt{RMAX} = 3.5 arcsec, corresponding to a \texttt{NUMR} = 23 quadratic spacings for the likelihood marginalization.

\begin{figure}
\centering
\includegraphics[scale=0.285]{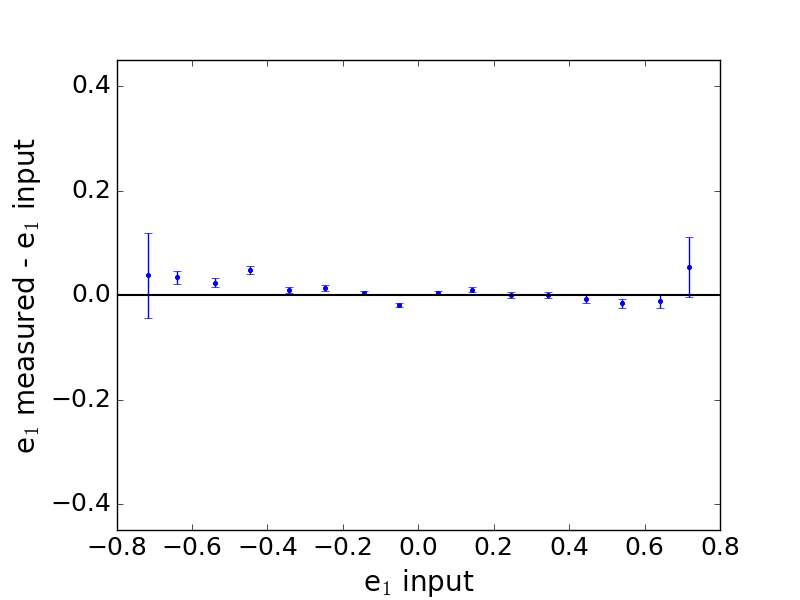}
\includegraphics[scale=0.285]{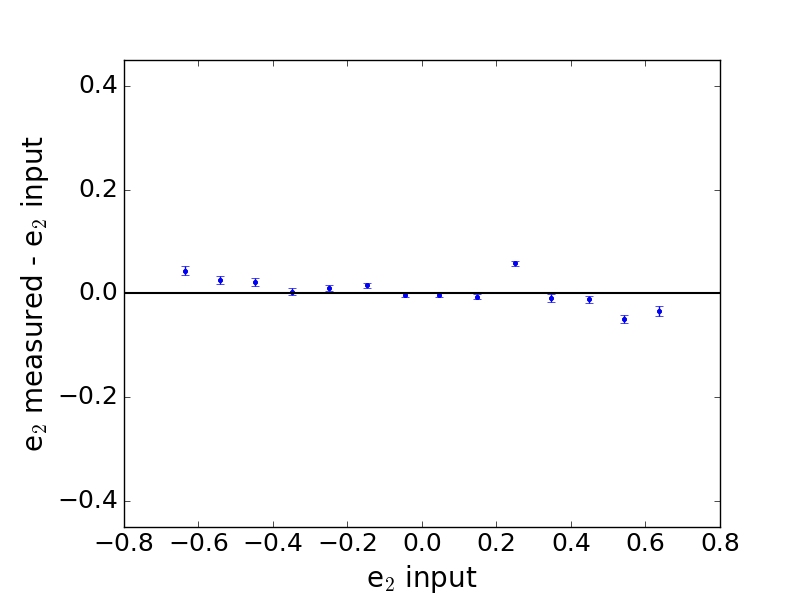}
\includegraphics[scale=0.285]{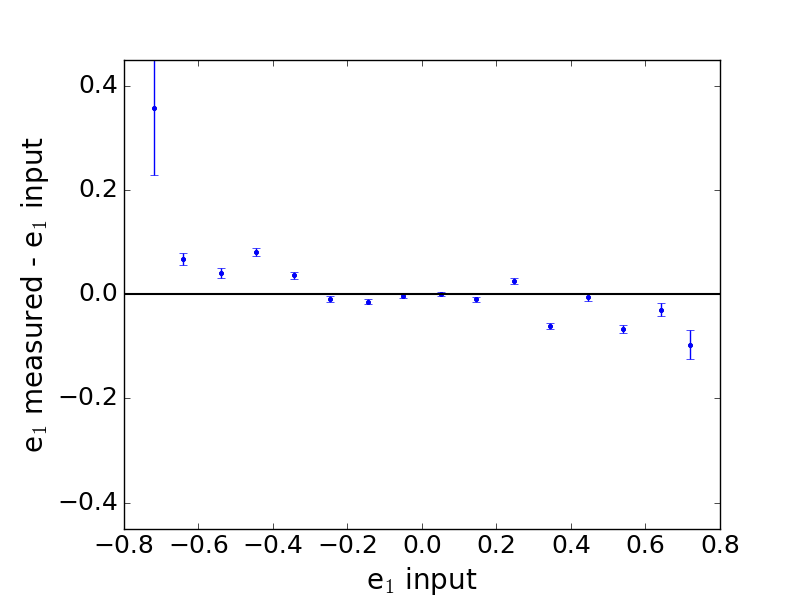}
\includegraphics[scale=0.285]{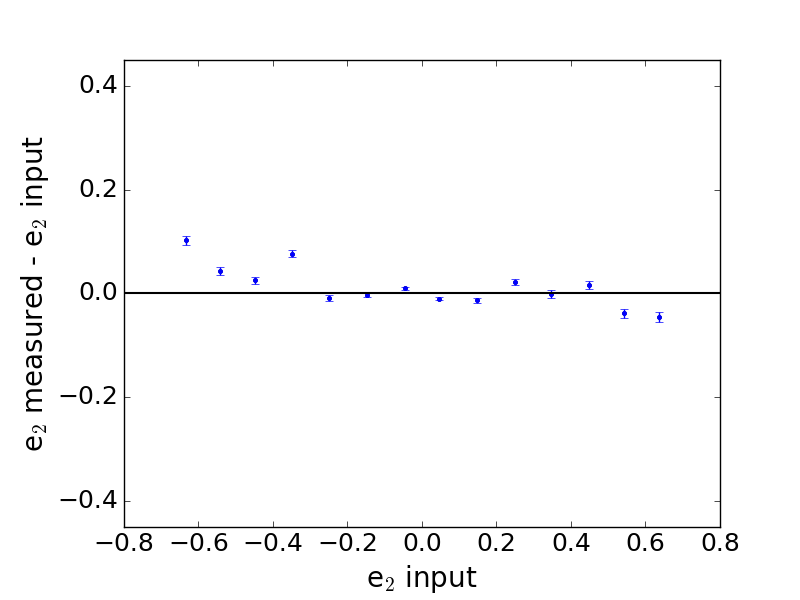}
\caption{Results for the SKA1-MID simulated dataset ($S \ge 10\mu$Jy): plots of the binned ''measured minus true'' values, $\Delta e_i$, for both source ellipticity components.  \textit{Upper panel}: one source at a time in the field of view. \textit{Lower panel}: all sources in the field of view at a density of 2.7~gal/armin$^2$.}
\label{fig:HMCdata}
\end{figure}

For the model fitting we adopt the optimal facet size factor selected in the previous section.  
As an intermediate step, we first measure ellipticities of a single source in the field of view, i.e. removing the other sources by simulating their visibilities with the real values of all the source parameters (taken  from the galaxy catalog of the simulation). In this case we have $1.3\%$ bad measurements, while in the case of the all sources in the field of view of 400 arcmin$^2$ the bad measurements are $1.4\%$. The plots of the binned measurements for the two cases are shown in Fig.~\ref{fig:HMCdata}. By comparing these results we see that our source extraction algorithm seems working efficiently. We also show that it can be improved in case the source sizes are known and used in the computation of the sky model. In particular, the number of bad sources reduces to $1\%$ reducing the shape bias at the same time. Table~\ref{tab:SKA1-MIDtest} contains the bias values of all these cases and the ones obtained in \cite{rivi2019} for comparison.

As expected, \textsc{RadioLensfit} shape measurements are less accurate than HMC, but they are comparable and we expect that the mean of the galaxy shapes obtained with the two methods may be closer when dealing with large statistics as for weak lensing shear measurement. On the other hand, the computation is much faster as shown in the next section.

\subsection{Computing time}
\label{sec:time}
Working on the same system  used for the HMC joint fitting (but without exploiting GPUs), i.e.  16-core Intel Xeon E5-2650 @ 2.00~GHz, we take a computational time of 1 h  35 min compared to more than 9 weeks taken with HMC and 2 NVIDIA Tesla K40 GPUs (see Table~3 of \cite{rivi2019}), which makes the HMC approach unfeasible with real observations, where there are thousands of frequency channels and a fine integration time.

\begin{table}
\scriptsize
\begin{tabular}{l c c c c c  }
\hline
Test & bad & \multicolumn{4}{c}{shape bias}  \\
  & & $m_1$ & $c_1$ & $m_2$ & $c_2$\\
\hline
RL single &  13 & $0.0322 \pm 0.0056$ & $0.0041 \pm 0.0015$  & $0.0321 \pm 0.0051$ & $0.0062 \pm 0.0015$ \\
RL all + sizes & 10 & $0.0402 \pm 0.0057$ & $0.0057 \pm 0.0016$  & $0.0393  \pm 0.0051$ & $0.0001\pm 0.0015$ \\
RL all &  14 & $0.0748 \pm 0.0056$ & $-0.0005 \pm 0.0015$  & $0.0620 \pm 0.0051$ & $0.0067 \pm 0.0015$ \\
HMC all &  0 & $0.0296 \pm 0.0043$ & $0.0001 \pm 0.0010$  & $0.0282 \pm 0.0040$ & $-0.0002 \pm 0.0010$ \\ 
\hline
\end{tabular}
\caption{Bias values for the fitting of the two ellipticity components of the SKA1-MID simulated observation (1000 sources in 400 arcmin$^2$  with $S \ge 10\mu$Jy) for the following  cases: a) \textsc{RadioLensfit}, single source in the field of view; b)  \textsc{RadioLensfit}, all sources in the field of view, extracted using the known size and position; c)  \textsc{RadioLensfit}, all sources in the field of view, extracted using the known flux and position; d) joint fitting of all sources with HMC (reported from \cite{rivi2019}).} 
\label{tab:SKA1-MIDtest}
\end{table}

\section{Scalability}
\label{sec:scalability}
 
Scalability tests are performed on the INAF cluster HOTCAT made of 20 nodes containing $4 \times 10$-core Intel Haswell E5-4627v3 @ 2.60~GHz with 256GB  DDR3 of shared memory, and a network Infiniband ConnectX 54~GB/sec. The parallel filesystem is based on a BeeGFS technology and the high performance storage (2GBs I/O) of capacity 600~TB is distributed over 4 nodes. The code is compiled with GNU/9.3.0 and OpenMPI/3.1.6.
Initially, we use the same simulated SKA1-MID observation of 1000 sources presented in Section~\ref{ska1-mid-sim}, where the uv coverage consists in 9,266,880 points and the IF band is made of a single channel. Then we re-simulate the same dataset but with more frequency channels and spectral windows, maintaining the overall frequency bandwidth.

\begin{figure}
\centering
\includegraphics[scale=0.285]{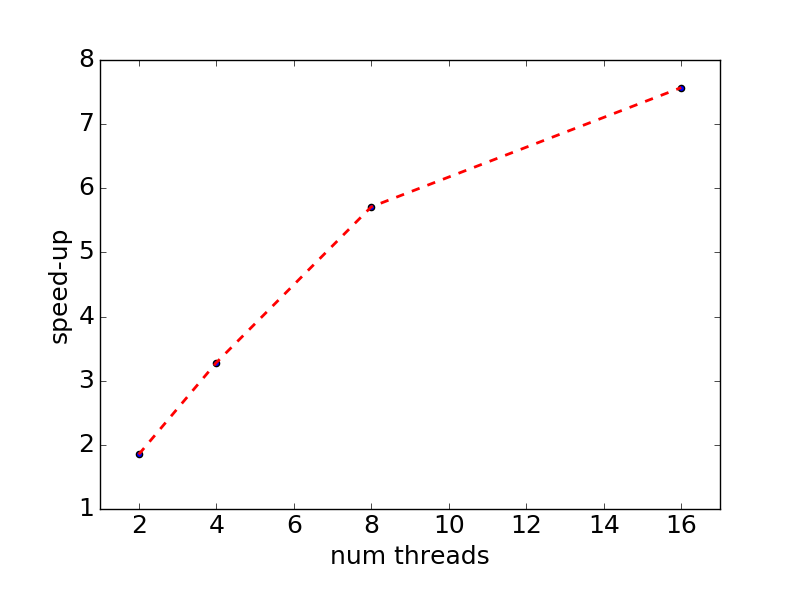}
\includegraphics[scale=0.285]{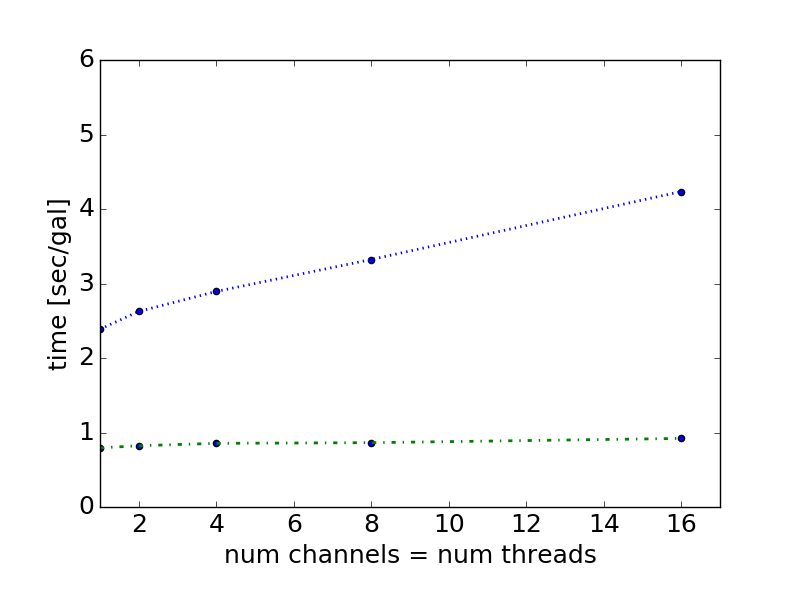}
\caption{OpenMP performance. \textit{Left panel:} Speed-up of the source model fitting, which is independent of the number of frequency channels. \textit{Right panel:} Plot of the computing times per galaxy for the sky model (green dashdot line) and source extraction (blue dotted line), varying the number of frequency channels (equal to the number of threads).}
\label{fig:speed-up}
\end{figure}

By running the serial version of the code we see that 3.5\% of the computing time is spent in the computation of the sky model, 10.9\% in the source extraction (including the sky model update) and the remaining 85.6\% in the source model fitting.
In the left panel of Fig.~\ref{fig:speed-up} we show the speed-up obtained with the OpenMP parallelization for the most intensive computational part of the code, i.e. the model fitting, whose parallelization is independent of the number of frequency channels in the spectral window. The speed-up is good up to 16 threads, where we reach a maximum of 7.6. It is not linear because of the sequential likelihood marginalization over the scalelength and threads creation overhead which occurs for each likelihood evaluation. Moreover the workload among threads is not exactly distributed when their number is not  a divider of \texttt{NUM\_R}. The right panel of Fig.~\ref{fig:speed-up} contains the plots of the computing time per galaxy for the sky model and source extraction, varying the number of frequency channels equal to the number of threads, and shows the full parallelization of the former over the number of channels and the residual effect of the serial faceting in the latter.  

\begin{figure}
\centering
\includegraphics[scale=0.45]{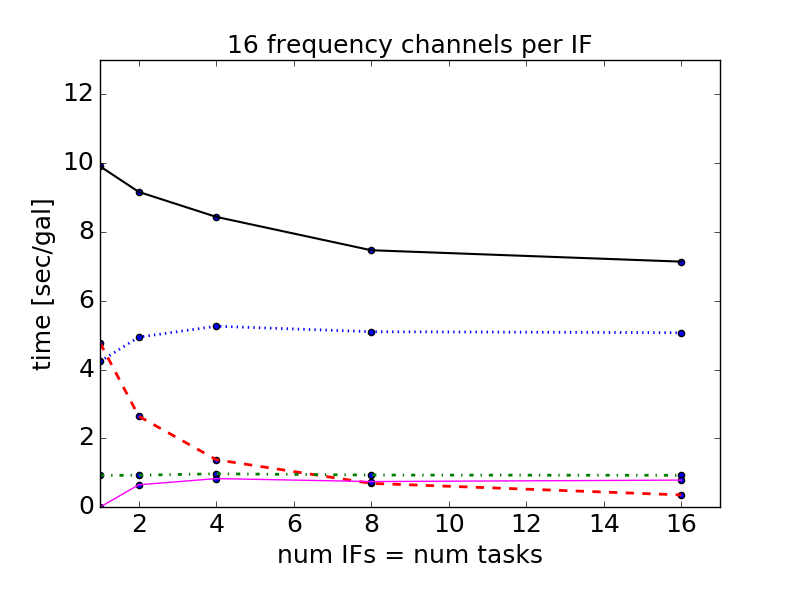}
\caption{MPI weak scalability (+16 OpenMP threads per task). Plot of the computing times per galaxy of the various phases of the computation, varying the number of spectral windows (equal to the number of MPI tasks): sky model (green dashdot line), source extraction (blue dotted line), model fitting (red dashed line), communication overhead (thin magenta solid line), total computing time (black solid line). Overall we obtain a super-linear weak scalability, despite tasks synchronization overhead for blocking communications, thanks to the strong scalability of the model fitting (sources are distributed among tasks).}
\label{fig:scalability}
\end{figure}

For the MPI version, we test the weak scalability, i.e. how the computing time varies with the number of tasks, each processing a dataset of the same size. We simulate IF bands made of 16 channels. For each test we double the number of spectral windows, so that each MPI task processes a single spectral window, and halve the channels bandwidth in order to maintain the total frequency bandwidth (and therefore source SNR). We perform the tests using 16 OpenMP threads per task and up to 16 MPI tasks. Fig.~\ref{fig:scalability} shows the plots of the processing time per source for each phase of the computation. As expected we reach a linear scalability for the sky model computation and source extraction as each process executes the same operations independently of the total number of spectral windows. The computing time for model fitting scales with the number of tasks as the number of sources to be processed per task reduces, while the number of facet visibilities is almost the same\footnote{Raw visibilities are averaged in a single facet with uv points measured in wavelength units.}. In the MPI communications we observe an overhead dominated by the tasks synchronization, which is due to the model fitting unbalance that may occur among sources as discussed in Section~\ref{sec:MPI}. In fact the real communication times are negligible: for 16 tasks we obtain a total time of about 15.5~sec. This overhead seems not changing with the number of processes probably because the unbalance in the model fitting is compensated by the reduction of the number of sources assigned to each task. 
Overall, we obtain a super-linear MPI weak scalability thanks to the model fitting strong scalability that largely compensates the communication overhead. 

\section{Towards real data analysis}
\label{sec:real_data}

The parallelization of the code, enabling the usage of multi-node systems, solves one of the main problems for the analysis of real data: the enormous size of raw visibilities. Working with real observations also requires the calibration of the visibilities for direction independent effects, e.g. antenna gains, and direction dependent effects (DDEs) such as the primary beam attenuation, bandwidth and time-average smearing. 
The primary beam pattern introduces a variation in sensitivity in the radial direction across the observing field of view which modifies the observed shape of source images: the antenna power pattern of the primary single dish $A(l, m)$ is multiplied by the surface brightness $I(l,m)$ before signals are correlated and is closely related to the diffraction pattern of a circular dish. 
Currently no instrumental systematics effects are added in the sky model but DDEs can be  modelled analytically \citep{Chang04,SmirnovA} and added in the visibilities computation when working with real data. For the treatment of these effects in the analysis of VLA FIRST Radio Survey see \cite{Chang04}.  

\begin{figure}
\centering
\includegraphics[scale=0.58]{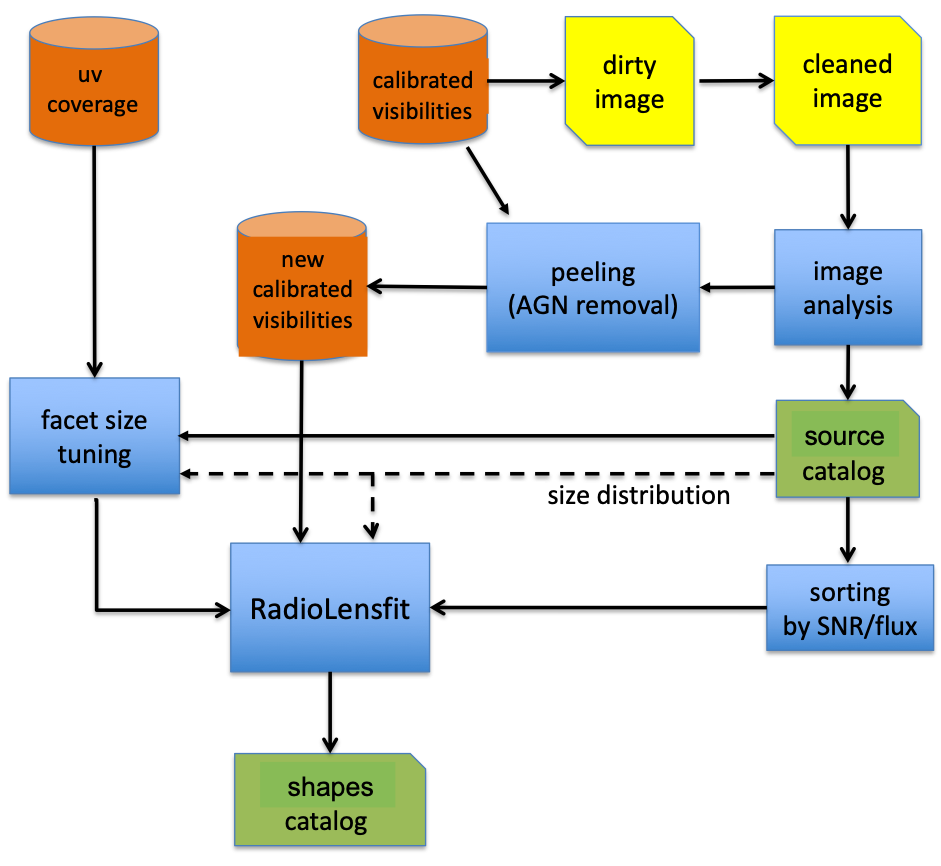}
\caption{Pipeline for radio weak lensing data analysis with \textsc{RadioLensfit}.}
\label{fig:pipeline}
\end{figure}

The imaging of the data is still necessary for peeling brightest sources, such as radio-loud AGN, and the measurement of sources position and flux, and possibly estimate size and shape, to produce the source catalog. Finally, as discussed in Section~\ref{sec:facet}, simulations with the real uv coverage have to be performed for tuning the facet size factor. Fig.~\ref{fig:pipeline} shows a possible pipeline for the usage of \textsc{RadioLensfit} on real data.

 Currently there is a work in progress within the SuperCLASS collaboration for the usage of \textsc{RadioLensfit} with the data of this precursor survey.
The weighting scheme for the isotropization of the PSF is still an open problem, but in the SuperCLASS survey the pointing centers and the integration time of the observation were chosen to make the PSF as isotropic as possibile. An investigation on shape models for RQ AGN could also be required as they may contaminate a sizeable fraction (20-30\%) of SF galaxies in the faint radio sky observed in radio weak lensing surveys \citep{Guidetti2017}. 

\section{Conclusions}
\label{sec:end}
We have presented \textsc{RadioLensfit}, a open-source scalable tool for an efficient and fast galaxy shape measurement for radio weak lensing. It works directly in the visibility domain, avoiding imaging systematics that may significantly affect the measurement. Single source visibilities are isolated and averaged in small facets in order to perform the model fitting of one galaxy at a time, as in the optical case. This approach enables the processing in a reasonable time of large numbers of sources and large sized datasets, compared with a more accurate but unrealistic joint fitting of all sources in the field of view (e.g. using HMC)\footnote{Note that HMC could possibly be applied in a second step for shape refinement of very clustered sources in small regions where a small number of sources could be simultaneously extracted for the joint fitting.}. 
This is shown for a simulated SKA-MID observation of 1000 star-forming galaxies at the expected source density in Phase~1 down to a source flux $S_{1.4\textrm{GHz}} = 10 \mu$Jy. Using a facet size source dependent, we obtain a multiplicative bias for the two source ellipticity components $m_1=0.0748 \pm 0.0056$ and $m_2 = 0.0620 \pm 0.0051$ that can be further improved if an estimate of the source size is provided in the source catalog. These bias values, comparable to the joint fitting case, lead to multiplicative shear bias about a factor 10 above the SKA1 Medium-Deep Band 2 Survey requirements \citep{Rivi16,Rivi18}. This is acceptable as for optical methods where multiplicative noise bias calibration is usually required by using simulations or a self-calibration approach, e.g. \cite{Zuntz18} and \cite{kannawadi19}.

We have implemented an hybrid parallelization MPI+OpenMP of the code to accelerate the computation by exploiting HPC infrastructures. The approach consists in splitting the dataset in spectral windows and distributing them to different MPI processes. Also sources are distributed among processes after joint source extraction and faceting. We tested MPI weak scalability using up to 16~nodes/256~cores of a Linux cluster, showing that the total computation scales more than linearly and with a good threading speed-up of the model fitting, which is the most computationally  intensive part of the code.  

This implementation allows to deal with observations where the spectral windows may have a different number of rows (uv points) and frequency channels, and to combine measurement of the same observation obtained by different interferometers (as for the precursor radio weak lensing survey SuperCLASS involving JVLA and e-MERLIN radio telescopes).

\section*{Acknowledgements}
MR acknowledges the support of the INAF Astronomical Observatory of Trieste, via the CHIPP pilot project \citep{CHIPP}, for the access to the HOTCAT cluster, and Filipe Batoni Abdalla for the access to the ''Splinter'' cluster at the Department of Physics and Astronomy of the University College London.

\bibliography{master}

\begin{thebibliography}{45}
\expandafter\ifx\csname natexlab\endcsname\relax\def\natexlab#1{#1}\fi
\providecommand{\url}[1]{\texttt{#1}}
\providecommand{\href}[2]{#2}
\providecommand{\path}[1]{#1}
\providecommand{\DOIprefix}{doi:}
\providecommand{\ArXivprefix}{arXiv:}
\providecommand{\URLprefix}{URL: }
\providecommand{\Pubmedprefix}{pmid:}
\providecommand{\doi}[1]{\href{http://dx.doi.org/#1}{\path{#1}}}
\providecommand{\Pubmed}[1]{\href{pmid:#1}{\path{#1}}}
\providecommand{\bibinfo}[2]{#2}
\ifx\xfnm\relax \def\xfnm[#1]{\unskip,\space#1}\fi
\bibitem[{Bacon et~al.(2020)Bacon, Battye, Bull, Camera, Ferreira, Harrison,
  Parkinson, Pourtsidou, Santos, Wolz, Abdalla, Akrami, Alonso, Andrianomena,
  Ballardini, Bernal, Bertacca, Bengaly, Bonaldi, Bonvin, Brown, Chapman, Chen,
  Chen, Cunnington, Davis, Dickinson, Fonseca, Grainge, Harper, Jarvis,
  Maartens, Maddox, Padmanabhan, Pritchard, Raccanelli, Rivi, Roychowdhury,
  Sahlen, Schwarz, Siewert, Viel, Villaescusa-Navarro, Xu, Yamauchi and
  Zuntz}]{RedBook2018}
\bibinfo{author}{Bacon, D.}, \bibinfo{author}{Battye, R.},
  \bibinfo{author}{Bull, P.}, \bibinfo{author}{Camera, S.},
  \bibinfo{author}{Ferreira, P.}, \bibinfo{author}{Harrison, I.},
  \bibinfo{author}{Parkinson, D.}, \bibinfo{author}{Pourtsidou, A.},
  \bibinfo{author}{Santos, M.}, \bibinfo{author}{Wolz, L.},
  \bibinfo{author}{Abdalla, F.}, \bibinfo{author}{Akrami, Y.},
  \bibinfo{author}{Alonso, D.}, \bibinfo{author}{Andrianomena, S.},
  \bibinfo{author}{Ballardini, M.}, \bibinfo{author}{Bernal, J.},
  \bibinfo{author}{Bertacca, D.}, \bibinfo{author}{Bengaly, C.},
  \bibinfo{author}{Bonaldi, A.}, \bibinfo{author}{Bonvin, C.},
  \bibinfo{author}{Brown, M.}, \bibinfo{author}{Chapman, E.},
  \bibinfo{author}{Chen, S.}, \bibinfo{author}{Chen, X.},
  \bibinfo{author}{Cunnington, S.}, \bibinfo{author}{Davis, T.},
  \bibinfo{author}{Dickinson, C.}, \bibinfo{author}{Fonseca, J.},
  \bibinfo{author}{Grainge, K.}, \bibinfo{author}{Harper, S.},
  \bibinfo{author}{Jarvis, M.}, \bibinfo{author}{Maartens, R.},
  \bibinfo{author}{Maddox, N.}, \bibinfo{author}{Padmanabhan, H.},
  \bibinfo{author}{Pritchard, J.}, \bibinfo{author}{Raccanelli, A.},
  \bibinfo{author}{Rivi, M.}, \bibinfo{author}{Roychowdhury, S.},
  \bibinfo{author}{Sahlen, M.}, \bibinfo{author}{Schwarz, D.},
  \bibinfo{author}{Siewert, T.}, \bibinfo{author}{Viel, M.},
  \bibinfo{author}{Villaescusa-Navarro, F.}, \bibinfo{author}{Xu, Y.},
  \bibinfo{author}{Yamauchi, D.}, \bibinfo{author}{Zuntz, J.},
  \bibinfo{year}{2020}.
\newblock \bibinfo{title}{{Cosmology with phase 1 of the Square Kilometre
  Array}}.
\newblock \bibinfo{journal}{Publ. Astron. Soc. Austral.} \bibinfo{volume}{37},
  \bibinfo{pages}{e007}.
\newblock \DOIprefix\doi{10.1017/pasa.2019.51}.
\bibitem[{Bonzini et~al.(2013)Bonzini, Padovani, Mainieri, Kellermann, Miller,
  Rosati, Tozzi and Vattakunnel}]{Bonzini2013}
\bibinfo{author}{Bonzini, M.}, \bibinfo{author}{Padovani, P.},
  \bibinfo{author}{Mainieri, V.}, \bibinfo{author}{Kellermann, K.I.},
  \bibinfo{author}{Miller, N.}, \bibinfo{author}{Rosati, P.},
  \bibinfo{author}{Tozzi, P.}, \bibinfo{author}{Vattakunnel, S.},
  \bibinfo{year}{2013}.
\newblock \bibinfo{title}{{The sub-mJy radio sky in the Extended Chandra Deep
  Field-South: source population}}.
\newblock \bibinfo{journal}{MNRAS} \bibinfo{volume}{436},
  \bibinfo{pages}{3759}.
\newblock \DOIprefix\doi{10.1093/mnras/stt1879}.
\bibitem[{{Briggs} et~al.(1999){Briggs}, {Schwab} and {Sramek}}]{Weighting99}
\bibinfo{author}{{Briggs}, D.}, \bibinfo{author}{{Schwab}, F.},
  \bibinfo{author}{{Sramek}, R.}, \bibinfo{year}{1999}.
\newblock \bibinfo{title}{Imaging}, in: \bibinfo{editor}{{Taylor}, G.},
  \bibinfo{editor}{{Carilli}, C.}, \bibinfo{editor}{{Perley}, R.} (Eds.),
  \bibinfo{booktitle}{Synthesis Imaging in Radio Astronomy II}.
  \bibinfo{publisher}{Astron. Soc. Pac., San Francisco}. volume
  \bibinfo{volume}{180} of \textit{\bibinfo{series}{ASP Conf. Ser.}}, p.
  \bibinfo{pages}{138}.
\bibitem[{Brown et~al.(2015)Brown, Bacon, Camera, Harrison, Joachimi, Metcalf,
  Pourtsidou, Takahashi, Zuntz, Abdalla, Bridle, Jarvis, Kitching, Miller and
  Patel}]{Brown15}
\bibinfo{author}{Brown, M.}, \bibinfo{author}{Bacon, D.},
  \bibinfo{author}{Camera, S.}, \bibinfo{author}{Harrison, I.},
  \bibinfo{author}{Joachimi, B.}, \bibinfo{author}{Metcalf, R.},
  \bibinfo{author}{Pourtsidou, A.}, \bibinfo{author}{Takahashi, K.},
  \bibinfo{author}{Zuntz, J.}, \bibinfo{author}{Abdalla, F.},
  \bibinfo{author}{Bridle, S.}, \bibinfo{author}{Jarvis, M.},
  \bibinfo{author}{Kitching, T.}, \bibinfo{author}{Miller, L.},
  \bibinfo{author}{Patel, P.}, \bibinfo{year}{2015}.
\newblock \bibinfo{title}{{Weak gravitational lensing with the Square Kilometre
  Array}}, in: \bibinfo{booktitle}{Advancing Astrophysics with the Square
  Kilometre Array}, \bibinfo{publisher}{Proc. Sci., SISSA, Trieste}. p.
  \bibinfo{pages}{PoS(AASKA14)023}.
\newblock \DOIprefix\doi{10.22323/1.215.0023}.
\bibitem[{{Brown} and {Battye}(2011)}]{BB11}
\bibinfo{author}{{Brown}, M.}, \bibinfo{author}{{Battye}, R.},
  \bibinfo{year}{2011}.
\newblock \bibinfo{title}{Polarization as an indicator of intrinsic alignment
  in radio weak lensing}.
\newblock \bibinfo{journal}{MNRAS} \bibinfo{volume}{410},
  \bibinfo{pages}{2057--2074}.
\newblock \DOIprefix\doi{10.1111/j.1365-2966.2010.17583.x}.
\bibitem[{{Camera} et~al.(2017){Camera}, {Harrison}, {Bonaldi} and
  {Brown}}]{Camera17}
\bibinfo{author}{{Camera}, S.}, \bibinfo{author}{{Harrison}, I.},
  \bibinfo{author}{{Bonaldi}, A.}, \bibinfo{author}{{Brown}, M.},
  \bibinfo{year}{2017}.
\newblock \bibinfo{title}{{SKA weak lensing – III. Added value of
  multiwavelength synergies for the mitigation of systematics}}.
\newblock \bibinfo{journal}{MNRAS} \bibinfo{volume}{464},
  \bibinfo{pages}{4747}.
\bibitem[{{Chang} and {Refregier}(2002)}]{CR02}
\bibinfo{author}{{Chang}, T.}, \bibinfo{author}{{Refregier}, A.},
  \bibinfo{year}{2002}.
\newblock \bibinfo{title}{Shape reconstruction and weak lensing measurement
  with interferometers: a shapelet approach}.
\newblock \bibinfo{journal}{ApJ} \bibinfo{volume}{570},
  \bibinfo{pages}{447--810}.
\newblock \DOIprefix\doi{10.1086/339496}.
\bibitem[{{Chang} et~al.(2004){Chang}, {Refregier} and {Helfand}}]{Chang04}
\bibinfo{author}{{Chang}, T.}, \bibinfo{author}{{Refregier}, A.},
  \bibinfo{author}{{Helfand}, D.}, \bibinfo{year}{2004}.
\newblock \bibinfo{title}{Weak lensing by large-scale structure with the
  {FIRST} radio survey}.
\newblock \bibinfo{journal}{ApJ} \bibinfo{volume}{617},
  \bibinfo{pages}{794--810}.
\newblock \DOIprefix\doi{10.1086/425491}.
\bibitem[{Delvecchio et~al.(2017)Delvecchio, Smolcic, Zamorani, Lagos, Berta,
  Delhaize, Baran, Alexander, Rosario, Gonzalez-Perez, Ilbert, Lacey, {Le
  Fevre}, Miettinen, Aravena, Bondi, Carilli, Ciliegi, Mooley, Novak,
  Schinnerer, Capak, Civano, Fanidakis, Ruiz, Karim, Laigle, Marchesi,
  McCracken, Middleberg, Salvato and Tasca}]{Delvecchio2017}
\bibinfo{author}{Delvecchio, I.}, \bibinfo{author}{Smolcic, V.},
  \bibinfo{author}{Zamorani, G.}, \bibinfo{author}{Lagos, C.D.P.},
  \bibinfo{author}{Berta, S.}, \bibinfo{author}{Delhaize, J.},
  \bibinfo{author}{Baran, N.}, \bibinfo{author}{Alexander, D.},
  \bibinfo{author}{Rosario, D.J.}, \bibinfo{author}{Gonzalez-Perez, V.},
  \bibinfo{author}{Ilbert, O.}, \bibinfo{author}{Lacey, C.G.},
  \bibinfo{author}{{Le Fevre}, O.}, \bibinfo{author}{Miettinen, O.},
  \bibinfo{author}{Aravena, M.}, \bibinfo{author}{Bondi, M.},
  \bibinfo{author}{Carilli, C.}, \bibinfo{author}{Ciliegi, P.},
  \bibinfo{author}{Mooley, K.}, \bibinfo{author}{Novak, M.},
  \bibinfo{author}{Schinnerer, E.}, \bibinfo{author}{Capak, P.},
  \bibinfo{author}{Civano, F.}, \bibinfo{author}{Fanidakis, N.},
  \bibinfo{author}{Ruiz, N.H.}, \bibinfo{author}{Karim, A.},
  \bibinfo{author}{Laigle, C.}, \bibinfo{author}{Marchesi, S.},
  \bibinfo{author}{McCracken, H.J.}, \bibinfo{author}{Middleberg, E.},
  \bibinfo{author}{Salvato, M.}, \bibinfo{author}{Tasca, L.},
  \bibinfo{year}{2017}.
\newblock \bibinfo{title}{The {VLA-COSMOS 3 GHz Large Project: AGN} and
  host-galaxy properties out to z $\le$ 6}.
\newblock \bibinfo{journal}{A\&A} \bibinfo{volume}{602}, \bibinfo{pages}{A3}.
\newblock \DOIprefix\doi{10.1051/0004-6361/201629367}.
\bibitem[{{Demetroullas} and {Brown}(2016)}]{DB16}
\bibinfo{author}{{Demetroullas}, C.}, \bibinfo{author}{{Brown}, M.},
  \bibinfo{year}{2016}.
\newblock \bibinfo{title}{{Cross-correlation cosmic shear with the SDSS and VLA
  FIRST surveys}}.
\newblock \bibinfo{journal}{MNRAS} \bibinfo{volume}{456},
  \bibinfo{pages}{3100--3118}.
\newblock \DOIprefix\doi{10.1093/mnras/stv2876}.
\bibitem[{Fu et~al.(2018)Fu, Liu, Radovich, Liu, Pan, Fan, Covone, Vaccari,
  Amaro, Brescia, Capaccioli, De~Cicco, Grado, Limatola, Miller, Napolitano,
  Paolillo and Pignata}]{Fu18}
\bibinfo{author}{Fu, L.}, \bibinfo{author}{Liu, D.}, \bibinfo{author}{Radovich,
  M.}, \bibinfo{author}{Liu, X.}, \bibinfo{author}{Pan, C.},
  \bibinfo{author}{Fan, Z.}, \bibinfo{author}{Covone, G.},
  \bibinfo{author}{Vaccari, M.}, \bibinfo{author}{Amaro, V.},
  \bibinfo{author}{Brescia, M.}, \bibinfo{author}{Capaccioli, M.},
  \bibinfo{author}{De~Cicco, D.}, \bibinfo{author}{Grado, A.},
  \bibinfo{author}{Limatola, L.}, \bibinfo{author}{Miller, L.},
  \bibinfo{author}{Napolitano, N.R.}, \bibinfo{author}{Paolillo, M.},
  \bibinfo{author}{Pignata, G.}, \bibinfo{year}{2018}.
\newblock \bibinfo{title}{Weak-lensing study in voice survey – i. shear
  measurement}.
\newblock \bibinfo{journal}{MNRAS} \bibinfo{volume}{479},
  \bibinfo{pages}{3858–3872}.
\newblock \DOIprefix\doi{10.1093/mnras/sty1579}.
\bibitem[{Giblin et~al.(2021)Giblin, Heymans, Asgari, Hildebrandt, Hoekstra,
  Joachimi, Kannawadi, Kuijken, Lin, Miller, Tröster, van~den Busch, Wright,
  Bilicki, Blake, de~Jong, Dvornik, Erben, Getman, Napolitano, Schneider, Shan
  and Valentijn}]{Giblin21}
\bibinfo{author}{Giblin, B.}, \bibinfo{author}{Heymans, C.},
  \bibinfo{author}{Asgari, M.}, \bibinfo{author}{Hildebrandt, H.},
  \bibinfo{author}{Hoekstra, H.}, \bibinfo{author}{Joachimi, B.},
  \bibinfo{author}{Kannawadi, A.}, \bibinfo{author}{Kuijken, K.},
  \bibinfo{author}{Lin, C.A.}, \bibinfo{author}{Miller, L.},
  \bibinfo{author}{Tröster, T.}, \bibinfo{author}{van~den Busch, J.L.},
  \bibinfo{author}{Wright, A.H.}, \bibinfo{author}{Bilicki, M.},
  \bibinfo{author}{Blake, C.}, \bibinfo{author}{de~Jong, J.},
  \bibinfo{author}{Dvornik, A.}, \bibinfo{author}{Erben, T.},
  \bibinfo{author}{Getman, F.}, \bibinfo{author}{Napolitano, N.R.},
  \bibinfo{author}{Schneider, P.}, \bibinfo{author}{Shan, H.},
  \bibinfo{author}{Valentijn, E.}, \bibinfo{year}{2021}.
\newblock \bibinfo{title}{Kids-1000 catalogue: Weak gravitational lensing shear
  measurements}.
\newblock \bibinfo{journal}{A\&A} \bibinfo{volume}{645}, \bibinfo{pages}{A105}.
\newblock \DOIprefix\doi{10.1051/0004-6361/202038850}.
\bibitem[{Guidetti et~al.(2017)Guidetti, Bondi, Prandoni, Muxlow, Beswick,
  Wrigley, Smail, McHardy, Thomson, Radcliffe and Argo}]{Guidetti2017}
\bibinfo{author}{Guidetti, D.}, \bibinfo{author}{Bondi, M.},
  \bibinfo{author}{Prandoni, I.}, \bibinfo{author}{Muxlow, T.},
  \bibinfo{author}{Beswick, R.}, \bibinfo{author}{Wrigley, N.},
  \bibinfo{author}{Smail, I.}, \bibinfo{author}{McHardy, I.},
  \bibinfo{author}{Thomson, A.P.}, \bibinfo{author}{Radcliffe, J.},
  \bibinfo{author}{Argo, M.K.}, \bibinfo{year}{2017}.
\newblock \bibinfo{title}{{The eMERGE Survey - I: Very Large Array 5.5 GHz
  observations of the GOODS-North Field}}.
\newblock \bibinfo{journal}{MNRAS} \bibinfo{volume}{471},
  \bibinfo{pages}{210--226}.
\newblock \DOIprefix\doi{10.1093/mnras/stx1162}.
\bibitem[{{Harrison} and {Brown}(2015)}]{SKA-ECP}
\bibinfo{author}{{Harrison}, I.}, \bibinfo{author}{{Brown}, M.},
  \bibinfo{year}{2015}.
\newblock \bibinfo{title}{{SKA Engineering Change Proposal: Gridded
  visibilities to enable precision cosmology with radio weak lensing}}.
\newblock \bibinfo{howpublished}{preprint (arXiv:1507.06639)}.
\bibitem[{Harrison et~al.(2020)Harrison, Brown, Tunbridge, Thomas, Hillier,
  Thomson, Whittaker, Abdalla, Battye, Bonaldi, Camera, Casey, Demetroullas,
  Hales, Jackson, Kay, Manning, Peters, Riseley and Watson}]{SuperCLASS-III}
\bibinfo{author}{Harrison, I.}, \bibinfo{author}{Brown, M.},
  \bibinfo{author}{Tunbridge, B.}, \bibinfo{author}{Thomas, D.},
  \bibinfo{author}{Hillier, T.}, \bibinfo{author}{Thomson, A.},
  \bibinfo{author}{Whittaker, L.}, \bibinfo{author}{Abdalla, F.},
  \bibinfo{author}{Battye, R.}, \bibinfo{author}{Bonaldi, A.},
  \bibinfo{author}{Camera, S.}, \bibinfo{author}{Casey, C.},
  \bibinfo{author}{Demetroullas, C.}, \bibinfo{author}{Hales, C.},
  \bibinfo{author}{Jackson, N.}, \bibinfo{author}{Kay, S.},
  \bibinfo{author}{Manning, S.}, \bibinfo{author}{Peters, A.},
  \bibinfo{author}{Riseley, C.}, \bibinfo{author}{Watson, R.S.C.},
  \bibinfo{year}{2020}.
\newblock \bibinfo{title}{{SuperCLASS – III. Weak lensing from radio and
  optical observations in Data Release 1}}.
\newblock \bibinfo{journal}{MNRAS} \bibinfo{volume}{495},
  \bibinfo{pages}{1737--1759}.
\newblock \DOIprefix\doi{10.1093/mnras/stw2688}.
\bibitem[{{Harrison} et~al.(2016){Harrison}, {Camera}, {Zuntz} and
  {Brown}}]{Harrison16}
\bibinfo{author}{{Harrison}, I.}, \bibinfo{author}{{Camera}, S.},
  \bibinfo{author}{{Zuntz}, J.}, \bibinfo{author}{{Brown}, M.},
  \bibinfo{year}{2016}.
\newblock \bibinfo{title}{{SKA weak lensing I: Cosmological forecasts and the
  power of radio-optical cross-correlations}}.
\newblock \bibinfo{journal}{MNRAS} \bibinfo{volume}{463},
  \bibinfo{pages}{3674}.
\newblock \DOIprefix\doi{10.1093/mnras/stw2082}.
\bibitem[{Heymans et~al.(2012)Heymans, Van~Waerbeke, Miller, Erben,
  Hildebrandt, Hoekstra, Kitching, Mellier, Simon, Bonnett, Coupon, Fu,
  Harnois-Déraps, Hudson, Kilbinger, Kuijken, Rowe, Schrabback, Semboloni, van
  Uitert, Vafaei and Velander}]{Heymans12}
\bibinfo{author}{Heymans, C.}, \bibinfo{author}{Van~Waerbeke, L.},
  \bibinfo{author}{Miller, L.}, \bibinfo{author}{Erben, T.},
  \bibinfo{author}{Hildebrandt, H.}, \bibinfo{author}{Hoekstra, H.},
  \bibinfo{author}{Kitching, T.D.}, \bibinfo{author}{Mellier, Y.},
  \bibinfo{author}{Simon, P.}, \bibinfo{author}{Bonnett, C.},
  \bibinfo{author}{Coupon, J.}, \bibinfo{author}{Fu, L.},
  \bibinfo{author}{Harnois-Déraps, J.}, \bibinfo{author}{Hudson, M.J.},
  \bibinfo{author}{Kilbinger, M.}, \bibinfo{author}{Kuijken, K.},
  \bibinfo{author}{Rowe, B.}, \bibinfo{author}{Schrabback, T.},
  \bibinfo{author}{Semboloni, E.}, \bibinfo{author}{van Uitert, E.},
  \bibinfo{author}{Vafaei, S.}, \bibinfo{author}{Velander, M.},
  \bibinfo{year}{2012}.
\newblock \bibinfo{title}{Cfhtlens: the canada–france–hawaii telescope
  lensing survey}.
\newblock \bibinfo{journal}{Monthly Notices of the Royal Astronomical Society}
  \bibinfo{volume}{427}, \bibinfo{pages}{146–166}.
\newblock \DOIprefix\doi{10.1111/j.1365-2966.2012.21952.x}.
\bibitem[{Hildebrandt et~al.(2016)Hildebrandt, Choi, Heymans, Blake, Erben,
  Miller, Nakajima, van Waerbeke, Viola, Buddendiek, Harnois-Déraps, Hojjati,
  Joachimi, Joudaki, Kitching, Wolf, Gwyn, Johnson, Kuijken, Sheikhbahaee,
  Tudorica and Yee}]{Hildebrandt16a}
\bibinfo{author}{Hildebrandt, H.}, \bibinfo{author}{Choi, A.},
  \bibinfo{author}{Heymans, C.}, \bibinfo{author}{Blake, C.},
  \bibinfo{author}{Erben, T.}, \bibinfo{author}{Miller, L.},
  \bibinfo{author}{Nakajima, R.}, \bibinfo{author}{van Waerbeke, L.},
  \bibinfo{author}{Viola, M.}, \bibinfo{author}{Buddendiek, A.},
  \bibinfo{author}{Harnois-Déraps, J.}, \bibinfo{author}{Hojjati, A.},
  \bibinfo{author}{Joachimi, B.}, \bibinfo{author}{Joudaki, S.},
  \bibinfo{author}{Kitching, T.D.}, \bibinfo{author}{Wolf, C.},
  \bibinfo{author}{Gwyn, S.}, \bibinfo{author}{Johnson, N.},
  \bibinfo{author}{Kuijken, K.}, \bibinfo{author}{Sheikhbahaee, Z.},
  \bibinfo{author}{Tudorica, A.}, \bibinfo{author}{Yee, H.K.C.},
  \bibinfo{year}{2016}.
\newblock \bibinfo{title}{Rcslens: The red cluster sequence lensing survey}.
\newblock \bibinfo{journal}{MNRAS} \bibinfo{volume}{463},
  \bibinfo{pages}{635–654}.
\newblock \DOIprefix\doi{10.1093/mnras/stw2013}.
\bibitem[{{Huff} et~al.(2019){Huff}, {Krause}, {Eifler}, {Fang}, {George} and
  {Schlegel}}]{HKEGS19}
\bibinfo{author}{{Huff}, E.}, \bibinfo{author}{{Krause}, E.},
  \bibinfo{author}{{Eifler}, T.}, \bibinfo{author}{{Fang}, X.},
  \bibinfo{author}{{George}, M.}, \bibinfo{author}{{Schlegel}, D.},
  \bibinfo{year}{2019}.
\newblock \bibinfo{title}{{Cosmic shear without shape noise}}.
\newblock \bibinfo{journal}{preprint (arXiv:1311.1489)} .
\bibitem[{Kannawadi et~al.(2019)Kannawadi, Hoekstra, Miller, Viola, Conti,
  Herbonnet, Erben, Heymans, Hildebrandt, Kuijken, Vakili and
  Wright}]{kannawadi19}
\bibinfo{author}{Kannawadi, A.}, \bibinfo{author}{Hoekstra, H.},
  \bibinfo{author}{Miller, L.}, \bibinfo{author}{Viola, M.},
  \bibinfo{author}{Conti, I.}, \bibinfo{author}{Herbonnet, R.},
  \bibinfo{author}{Erben, T.}, \bibinfo{author}{Heymans, C.},
  \bibinfo{author}{Hildebrandt, H.}, \bibinfo{author}{Kuijken, K.},
  \bibinfo{author}{Vakili, M.}, \bibinfo{author}{Wright, A.},
  \bibinfo{year}{2019}.
\newblock \bibinfo{title}{{Towards emulating cosmic shear data: revisiting the
  calibration of the shear measurements for the Kilo-Degree Survey}}.
\newblock \bibinfo{journal}{A\&A} \bibinfo{volume}{624}, \bibinfo{pages}{A92}.
\newblock \DOIprefix\doi{10.1051/0004-6361/201834819}.
\bibitem[{{Kilbinger}(2015)}]{Kil15}
\bibinfo{author}{{Kilbinger}, M.}, \bibinfo{year}{2015}.
\newblock \bibinfo{title}{Cosmology with cosmic shear observations: a review}.
\newblock \bibinfo{journal}{Rep. Prog. Phys.} \bibinfo{volume}{78},
  \bibinfo{pages}{086901}.
\newblock \DOIprefix\doi{10.1088/0034-4885/78/8/086901}.
\bibitem[{Kuijken et~al.(2015)Kuijken, Heymans, Hildebrandt, Nakajima, Erben,
  de~Jong, Viola, Choi, Hoekstra, Miller, van Uitert, Amon, Blake, Brouwer,
  Buddendiek, Conti, Eriksen, Grado, Harnois-Déraps, Helmich, Herbonnet,
  Irisarri, Kitching, Klaes, La~Barbera, Napolitano, Radovich, Schneider,
  Sifón, Sikkema, Simon, Tudorica, Valentijn, Verdoes~Kleijn and van
  Waerbeke}]{Kuijken15}
\bibinfo{author}{Kuijken, K.}, \bibinfo{author}{Heymans, C.},
  \bibinfo{author}{Hildebrandt, H.}, \bibinfo{author}{Nakajima, R.},
  \bibinfo{author}{Erben, T.}, \bibinfo{author}{de~Jong, J.T.A.},
  \bibinfo{author}{Viola, M.}, \bibinfo{author}{Choi, A.},
  \bibinfo{author}{Hoekstra, H.}, \bibinfo{author}{Miller, L.},
  \bibinfo{author}{van Uitert, E.}, \bibinfo{author}{Amon, A.},
  \bibinfo{author}{Blake, C.}, \bibinfo{author}{Brouwer, M.},
  \bibinfo{author}{Buddendiek, A.}, \bibinfo{author}{Conti, I.F.},
  \bibinfo{author}{Eriksen, M.}, \bibinfo{author}{Grado, A.},
  \bibinfo{author}{Harnois-Déraps, J.}, \bibinfo{author}{Helmich, E.},
  \bibinfo{author}{Herbonnet, R.}, \bibinfo{author}{Irisarri, N.},
  \bibinfo{author}{Kitching, T.}, \bibinfo{author}{Klaes, D.},
  \bibinfo{author}{La~Barbera, F.}, \bibinfo{author}{Napolitano, N.},
  \bibinfo{author}{Radovich, M.}, \bibinfo{author}{Schneider, P.},
  \bibinfo{author}{Sifón, C.}, \bibinfo{author}{Sikkema, G.},
  \bibinfo{author}{Simon, P.}, \bibinfo{author}{Tudorica, A.},
  \bibinfo{author}{Valentijn, E.}, \bibinfo{author}{Verdoes~Kleijn, G.},
  \bibinfo{author}{van Waerbeke, L.}, \bibinfo{year}{2015}.
\newblock \bibinfo{title}{Gravitational lensing analysis of the kilo-degree
  survey}.
\newblock \bibinfo{journal}{MNRAS} \bibinfo{volume}{454},
  \bibinfo{pages}{3500–3532}.
\newblock \DOIprefix\doi{10.1093/mnras/stv2140}.
\bibitem[{Mancuso et~al.(2017)Mancuso, Lapi, Prandoni, Obi, Gonzalez-Nuevo,
  Perrotta, Bressan, Celotti and Danese}]{Mancuso2017}
\bibinfo{author}{Mancuso, C.}, \bibinfo{author}{Lapi, A.},
  \bibinfo{author}{Prandoni, I.}, \bibinfo{author}{Obi, I.},
  \bibinfo{author}{Gonzalez-Nuevo, J.}, \bibinfo{author}{Perrotta, F.},
  \bibinfo{author}{Bressan, A.}, \bibinfo{author}{Celotti, A.},
  \bibinfo{author}{Danese, L.}, \bibinfo{year}{2017}.
\newblock \bibinfo{title}{Galaxy evolution in the radio band: the role of
  star-forming galaxies and active galactic nuclei}.
\newblock \bibinfo{journal}{ApJ} \bibinfo{volume}{842}, \bibinfo{pages}{95}.
\newblock \DOIprefix\doi{10.3847/1538-4357/aa745d}.
\bibitem[{{Mandelbaum} et~al.(2018){Mandelbaum}, {Lanusse}, {Leauthaud},
  {Armstrong}, {Simet}, {Miyatake}, {Meyers}, {Bosch}, {Miyazaki} and
  {Tanaka}}]{mandelbaum18}
\bibinfo{author}{{Mandelbaum}, R.}, \bibinfo{author}{{Lanusse}, F.},
  \bibinfo{author}{{Leauthaud}, A.}, \bibinfo{author}{{Armstrong}, R.},
  \bibinfo{author}{{Simet}, M.}, \bibinfo{author}{{Miyatake}, H.},
  \bibinfo{author}{{Meyers}, J.E.}, \bibinfo{author}{{Bosch}, J.},
  \bibinfo{author}{{Miyazaki}, S.}, \bibinfo{author}{{Tanaka}, M.},
  \bibinfo{year}{2018}.
\newblock \bibinfo{title}{{Weak lensing shear calibration with simulations of
  the HSC survey}}.
\newblock \bibinfo{journal}{MNRAS} \bibinfo{volume}{481},
  \bibinfo{pages}{3170--3195}.
\newblock \DOIprefix\doi{10.1093/mnras/sty2420}.
\bibitem[{{Mandelbaum} et~al.(2015){Mandelbaum}, {Rowe}, {Armstrong}, {Bard},
  {Bertin}, {Bosch}, {Boutigny}, {Courbin}, {Dawson}, {Donnarumma}, {Fenech
  Conti}, {Gavazzi}, {Gentile}, {Gill}, {Hogg}, {Huff}, {Jee}, {Kacprzak},
  {Kilbinger}, {Kuntzer}, {Lang}, {Luo}, {March}, {Marshall}, {Meyers},
  {Miller}, {Miyatake}, {Nakajima}, {Ngol{\'e} Mboula}, {Nurbaeva}, {Okura},
  {Paulin-Henriksson}, {Rhodes}, {Schneider}, {Shan}, {Sheldon}, {Simet},
  {Starck}, {Sureau}, {Tewes}, {Zarb Adami}, {Zhang} and
  {Zuntz}}]{Mandelbaum15}
\bibinfo{author}{{Mandelbaum}, R.}, \bibinfo{author}{{Rowe}, B.},
  \bibinfo{author}{{Armstrong}, R.}, \bibinfo{author}{{Bard}, D.},
  \bibinfo{author}{{Bertin}, E.}, \bibinfo{author}{{Bosch}, J.},
  \bibinfo{author}{{Boutigny}, D.}, \bibinfo{author}{{Courbin}, F.},
  \bibinfo{author}{{Dawson}, W.A.}, \bibinfo{author}{{Donnarumma}, A.},
  \bibinfo{author}{{Fenech Conti}, I.}, \bibinfo{author}{{Gavazzi}, R.},
  \bibinfo{author}{{Gentile}, M.}, \bibinfo{author}{{Gill}, M.S.S.},
  \bibinfo{author}{{Hogg}, D.W.}, \bibinfo{author}{{Huff}, E.M.},
  \bibinfo{author}{{Jee}, M.J.}, \bibinfo{author}{{Kacprzak}, T.},
  \bibinfo{author}{{Kilbinger}, M.}, \bibinfo{author}{{Kuntzer}, T.},
  \bibinfo{author}{{Lang}, D.}, \bibinfo{author}{{Luo}, W.},
  \bibinfo{author}{{March}, M.C.}, \bibinfo{author}{{Marshall}, P.J.},
  \bibinfo{author}{{Meyers}, J.E.}, \bibinfo{author}{{Miller}, L.},
  \bibinfo{author}{{Miyatake}, H.}, \bibinfo{author}{{Nakajima}, R.},
  \bibinfo{author}{{Ngol{\'e} Mboula}, F.M.}, \bibinfo{author}{{Nurbaeva}, G.},
  \bibinfo{author}{{Okura}, Y.}, \bibinfo{author}{{Paulin-Henriksson}, S.},
  \bibinfo{author}{{Rhodes}, J.}, \bibinfo{author}{{Schneider}, M.D.},
  \bibinfo{author}{{Shan}, H.}, \bibinfo{author}{{Sheldon}, E.S.},
  \bibinfo{author}{{Simet}, M.}, \bibinfo{author}{{Starck}, J.L.},
  \bibinfo{author}{{Sureau}, F.}, \bibinfo{author}{{Tewes}, M.},
  \bibinfo{author}{{Zarb Adami}, K.}, \bibinfo{author}{{Zhang}, J.},
  \bibinfo{author}{{Zuntz}, J.}, \bibinfo{year}{2015}.
\newblock \bibinfo{title}{{GREAT3 results - I. Systematic errors in shear
  estimation and the impact of real galaxy morphology}}.
\newblock \bibinfo{journal}{MNRAS} \bibinfo{volume}{450},
  \bibinfo{pages}{2963--3007}.
\newblock \DOIprefix\doi{10.1093/mnras/stv781}.
\bibitem[{{Melchior} et~al.(2010){Melchior}, {B{\"o}hnert}, {Lombardi} and
  {Bartelmann}}]{Melchior10}
\bibinfo{author}{{Melchior}, P.}, \bibinfo{author}{{B{\"o}hnert}, A.},
  \bibinfo{author}{{Lombardi}, M.}, \bibinfo{author}{{Bartelmann}, M.},
  \bibinfo{year}{2010}.
\newblock \bibinfo{title}{{Limitations on shapelet based weak lensing
  measurements}}.
\newblock \bibinfo{journal}{A\&A} \bibinfo{volume}{510}, \bibinfo{pages}{A75}.
\bibitem[{{Miller} et~al.(2013){Miller}, {Heymans}, {Kitching}, {van Waerbeke},
  {Erben}, {Hildebrandt}, {Hoekstra}, {Mellier}, {Rowe}, {Coupon}, {Dietrich},
  {Fu}, {Harnois-D{\'e}raps}, {Hudson}, {Kilbinger}, {Kuijken}, {Schrabback},
  {Semboloni}, {Vafaei} and {Velander}}]{Miller13}
\bibinfo{author}{{Miller}, L.}, \bibinfo{author}{{Heymans}, C.},
  \bibinfo{author}{{Kitching}, T.D.}, \bibinfo{author}{{van Waerbeke}, L.},
  \bibinfo{author}{{Erben}, T.}, \bibinfo{author}{{Hildebrandt}, H.},
  \bibinfo{author}{{Hoekstra}, H.}, \bibinfo{author}{{Mellier}, Y.},
  \bibinfo{author}{{Rowe}, B.T.P.}, \bibinfo{author}{{Coupon}, J.},
  \bibinfo{author}{{Dietrich}, J.P.}, \bibinfo{author}{{Fu}, L.},
  \bibinfo{author}{{Harnois-D{\'e}raps}, J.}, \bibinfo{author}{{Hudson}, M.J.},
  \bibinfo{author}{{Kilbinger}, M.}, \bibinfo{author}{{Kuijken}, K.},
  \bibinfo{author}{{Schrabback}, T.}, \bibinfo{author}{{Semboloni}, E.},
  \bibinfo{author}{{Vafaei}, S.}, \bibinfo{author}{{Velander}, M.},
  \bibinfo{year}{2013}.
\newblock \bibinfo{title}{{Bayesian galaxy shape measurement for weak lensing
  surveys - III. Application to the Canada-France-Hawaii Telescope lensing
  survey}}.
\newblock \bibinfo{journal}{MNRAS} \bibinfo{volume}{429},
  \bibinfo{pages}{2858--2880}.
\newblock \DOIprefix\doi{10.1093/mnras/sts454}.
\bibitem[{Neal(2011)}]{Neal11}
\bibinfo{author}{Neal, R.M.}, \bibinfo{year}{2011}.
\newblock \bibinfo{title}{{MCMC using Hamiltonian dynamics}}, in:
  \bibinfo{booktitle}{Handbook of Markov Chain Monte Carlo}.
  \bibinfo{publisher}{Chapman \& Hall/CRC Press}, p. \bibinfo{pages}{113}.
\bibitem[{{Nelder} and {Mead}(1965)}]{Nelder-Mead}
\bibinfo{author}{{Nelder}, J.}, \bibinfo{author}{{Mead}, R.},
  \bibinfo{year}{1965}.
\newblock \bibinfo{title}{A simplex method for function minimization}.
\newblock \bibinfo{journal}{Comp. J.} \bibinfo{volume}{7},
  \bibinfo{pages}{308--313}.
\bibitem[{Owen(2018)}]{GOODSN2018}
\bibinfo{author}{Owen, F.}, \bibinfo{year}{2018}.
\newblock \bibinfo{title}{{Deep JVLA imaging of GOODS-N at 20cm}}.
\newblock \bibinfo{journal}{ApJ Supp. Series} \bibinfo{volume}{235},
  \bibinfo{pages}{34}.
\bibitem[{{Patel} et~al.(2010){Patel}, {Bacon}, {Beswick}, {Muxlow} and
  {Hoyle}}]{Patel10}
\bibinfo{author}{{Patel}, P.}, \bibinfo{author}{{Bacon}, D.J.},
  \bibinfo{author}{{Beswick}, R.J.}, \bibinfo{author}{{Muxlow}, T.W.B.},
  \bibinfo{author}{{Hoyle}, B.}, \bibinfo{year}{2010}.
\newblock \bibinfo{title}{Radio weak gravitational lensing with {VLA and
  MERLIN}}.
\newblock \bibinfo{journal}{MNRAS} \bibinfo{volume}{401},
  \bibinfo{pages}{2572--2586}.
\newblock \DOIprefix\doi{10.1111/j.1365-2966.2009.15836.x}.
\bibitem[{{Patel} et~al.(2015){Patel}, {Harrison}, {Makhathini}, {Abdalla},
  {Bacon}, {Brown}, {Heywood}, {Jarvis} and {Smirnov}}]{Patel15}
\bibinfo{author}{{Patel}, P.}, \bibinfo{author}{{Harrison}, I.},
  \bibinfo{author}{{Makhathini}, S.}, \bibinfo{author}{{Abdalla}, F.},
  \bibinfo{author}{{Bacon}, D.}, \bibinfo{author}{{Brown}, M.},
  \bibinfo{author}{{Heywood}, I.}, \bibinfo{author}{{Jarvis}, M.},
  \bibinfo{author}{{Smirnov}, O.}, \bibinfo{year}{2015}.
\newblock \bibinfo{title}{{Weak lensing simulations for the SKA}}, in:
  \bibinfo{booktitle}{Advancing Astrophysics with the Square Kilometre Array},
  \bibinfo{publisher}{Proc. Sci., SISSA, Trieste}. p.
  \bibinfo{pages}{PoS(AASKA14)030}.
\newblock \DOIprefix\doi{10.22323/1.215.0030}.
\bibitem[{{Refregier}(2003)}]{Refregier03}
\bibinfo{author}{{Refregier}, A.}, \bibinfo{year}{2003}.
\newblock \bibinfo{title}{{Shapelets – I. A method for image analysis}}.
\newblock \bibinfo{journal}{MNRAS} \bibinfo{volume}{338}, \bibinfo{pages}{35}.
\bibitem[{{Refregier} and {Bacon}(2003)}]{RB2003}
\bibinfo{author}{{Refregier}, A.}, \bibinfo{author}{{Bacon}, D.},
  \bibinfo{year}{2003}.
\newblock \bibinfo{title}{{Shapelets – II. A method for weak lensing
  measurements}}.
\newblock \bibinfo{journal}{MNRAS} \bibinfo{volume}{338}, \bibinfo{pages}{48}.
\newblock \DOIprefix\doi{10.1046/j.1365-8711.2003.05901.x}.
\bibitem[{Rivi et~al.(2019)Rivi, Lochner, Balan, Harrison and
  Abdalla}]{rivi2019}
\bibinfo{author}{Rivi, M.}, \bibinfo{author}{Lochner, M.},
  \bibinfo{author}{Balan, S.}, \bibinfo{author}{Harrison, I.},
  \bibinfo{author}{Abdalla, F.}, \bibinfo{year}{2019}.
\newblock \bibinfo{title}{{Radio galaxy shape measurement with Hamiltonian
  Monte Carlo in the visibility domain}}.
\newblock \bibinfo{journal}{MNRAS} \bibinfo{volume}{482},
  \bibinfo{pages}{1096--1109}.
\newblock \DOIprefix\doi{10.1093/mnras/sty2700}.
\bibitem[{{Rivi} and {Miller}(2018)}]{Rivi18}
\bibinfo{author}{{Rivi}, M.}, \bibinfo{author}{{Miller}, L.},
  \bibinfo{year}{2018}.
\newblock \bibinfo{title}{{Radio weak lensing shear measurement in the
  visibility domain - II. Source extraction}}.
\newblock \bibinfo{journal}{MNRAS} \bibinfo{volume}{476},
  \bibinfo{pages}{2053}.
\newblock \DOIprefix\doi{10.1093/mnras/sty371}.
\bibitem[{{Rivi} et~al.(2016a){Rivi}, {Miller}, {Makhathini} and
  {Abdalla}}]{Rivi15}
\bibinfo{author}{{Rivi}, M.}, \bibinfo{author}{{Miller}, L.},
  \bibinfo{author}{{Makhathini}, S.}, \bibinfo{author}{{Abdalla}, F.},
  \bibinfo{year}{2016}a.
\newblock \bibinfo{title}{Radiolensfit: Bayesian weak lensing measurement in
  the visibility domain}, in: \bibinfo{booktitle}{The many facets of
  extragalactic radio surveys: towards new scientific challenges},
  \bibinfo{publisher}{Proc. Sci., SISSA, Trieste}. pp.
  \bibinfo{pages}{PoS(EXTRA--RADSUR2015)033}.
\newblock \DOIprefix\doi{10.22323/1.267.0033}.
\bibitem[{{Rivi} et~al.(2016b){Rivi}, {Miller}, {Makhathini} and
  {Abdalla}}]{Rivi16}
\bibinfo{author}{{Rivi}, M.}, \bibinfo{author}{{Miller}, L.},
  \bibinfo{author}{{Makhathini}, S.}, \bibinfo{author}{{Abdalla}, F.B.},
  \bibinfo{year}{2016}b.
\newblock \bibinfo{title}{{Radio weak lensing shear measurement in the
  visibility domain - I. Methodology}}.
\newblock \bibinfo{journal}{MNRAS} \bibinfo{volume}{463},
  \bibinfo{pages}{1881}.
\newblock \DOIprefix\doi{10.1093/mnras/stw2041}.
\bibitem[{{Samuroff} et~al.(2018){Samuroff}, {Bridle}, {Zuntz}, {Troxel},
  {Gruen}, {Rollins}, {Bernstein}, {Eifler}, {Huff}, {Kacprzak}, {Krause},
  {MacCrann}, {Abdalla}, {Allam}, {Annis}, {Bechtol}, {Benoit-Levy}, {Bertin},
  {Brooks}, {Buckley-Geer}, {Carnero Rosell}, {Carrasco Kind}, {Carretero},
  {Crocce}, {D'Andrea}, {da Costa}, {Davis}, {Desai}, {Doel}, {Fausti Neto},
  {Flaugher}, {Fosalba}, {Frieman}, {Garcia-Bellido}, {Gerdes}, {Gruendl},
  {Gschwend}, {Gutierrez}, {Honscheid}, {James}, {Jarvis}, {Jeltema}, {Kirk},
  {Kuehn}, {Kuhlmann}, {Li}, {Lima}, {Maia}, {March}, {Marshall}, {Martini},
  {Melchior}, {Menanteau}, {Miquel}, {Nord}, {Ogando}, {Plazas}, {Roodman},
  {Sanchez}, {Scarpine}, {Schindler}, {Schubnell}, {Sevilla-Noarbe}, {Sheldon},
  {Smith}, {Soares-Santos}, {Sobreira}, {Suchyta}, {Tarle}, {Thomas} and
  {Tucker}}]{Samuroff18}
\bibinfo{author}{{Samuroff}, S.}, \bibinfo{author}{{Bridle}, S.L.},
  \bibinfo{author}{{Zuntz}, J.}, \bibinfo{author}{{Troxel}, M.A.},
  \bibinfo{author}{{Gruen}, D.}, \bibinfo{author}{{Rollins}, R.P.},
  \bibinfo{author}{{Bernstein}, G.M.}, \bibinfo{author}{{Eifler}, T.F.},
  \bibinfo{author}{{Huff}, E.M.}, \bibinfo{author}{{Kacprzak}, T.},
  \bibinfo{author}{{Krause}, E.}, \bibinfo{author}{{MacCrann}, N.},
  \bibinfo{author}{{Abdalla}, F.B.}, \bibinfo{author}{{Allam}, S.},
  \bibinfo{author}{{Annis}, J.}, \bibinfo{author}{{Bechtol}, K.},
  \bibinfo{author}{{Benoit-Levy}, A.}, \bibinfo{author}{{Bertin}, E.},
  \bibinfo{author}{{Brooks}, D.}, \bibinfo{author}{{Buckley-Geer}, E.},
  \bibinfo{author}{{Carnero Rosell}, A.}, \bibinfo{author}{{Carrasco Kind},
  M.}, \bibinfo{author}{{Carretero}, J.}, \bibinfo{author}{{Crocce}, M.},
  \bibinfo{author}{{D'Andrea}, C.B.}, \bibinfo{author}{{da Costa}, L.N.},
  \bibinfo{author}{{Davis}, C.}, \bibinfo{author}{{Desai}, S.},
  \bibinfo{author}{{Doel}, P.}, \bibinfo{author}{{Fausti Neto}, A.},
  \bibinfo{author}{{Flaugher}, B.}, \bibinfo{author}{{Fosalba}, P.},
  \bibinfo{author}{{Frieman}, J.}, \bibinfo{author}{{Garcia-Bellido}, J.},
  \bibinfo{author}{{Gerdes}, D.W.}, \bibinfo{author}{{Gruendl}, R.A.},
  \bibinfo{author}{{Gschwend}, J.}, \bibinfo{author}{{Gutierrez}, G.},
  \bibinfo{author}{{Honscheid}, K.}, \bibinfo{author}{{James}, D.J.},
  \bibinfo{author}{{Jarvis}, M.}, \bibinfo{author}{{Jeltema}, T.},
  \bibinfo{author}{{Kirk}, D.}, \bibinfo{author}{{Kuehn}, K.},
  \bibinfo{author}{{Kuhlmann}, S.}, \bibinfo{author}{{Li}, T.S.},
  \bibinfo{author}{{Lima}, M.}, \bibinfo{author}{{Maia}, M.A.G.},
  \bibinfo{author}{{March}, M.}, \bibinfo{author}{{Marshall}, J.L.},
  \bibinfo{author}{{Martini}, P.}, \bibinfo{author}{{Melchior}, P.},
  \bibinfo{author}{{Menanteau}, F.}, \bibinfo{author}{{Miquel}, R.},
  \bibinfo{author}{{Nord}, B.}, \bibinfo{author}{{Ogando}, R.L.C.},
  \bibinfo{author}{{Plazas}, A.A.}, \bibinfo{author}{{Roodman}, A.},
  \bibinfo{author}{{Sanchez}, E.}, \bibinfo{author}{{Scarpine}, V.},
  \bibinfo{author}{{Schindler}, R.}, \bibinfo{author}{{Schubnell}, M.},
  \bibinfo{author}{{Sevilla-Noarbe}, I.}, \bibinfo{author}{{Sheldon}, E.},
  \bibinfo{author}{{Smith}, M.}, \bibinfo{author}{{Soares-Santos}, M.},
  \bibinfo{author}{{Sobreira}, F.}, \bibinfo{author}{{Suchyta}, E.},
  \bibinfo{author}{{Tarle}, G.}, \bibinfo{author}{{Thomas}, D.},
  \bibinfo{author}{{Tucker}, D.L.}, \bibinfo{year}{2018}.
\newblock \bibinfo{title}{{Dark Energy Survey Year 1 Results: The Impact of
  Galaxy Neighbours on Weak Lensing Cosmology with IM3SHAPE}}.
\newblock \bibinfo{journal}{MNRAS} \bibinfo{volume}{475},
  \bibinfo{pages}{4524--4543}.
\newblock \DOIprefix\doi{10.1093/mnras/stx3282}.
\bibitem[{{Smirnov}(2011)}]{SmirnovA}
\bibinfo{author}{{Smirnov}, O.M.}, \bibinfo{year}{2011}.
\newblock \bibinfo{title}{{Revisiting the radio interferometer measurement
  equation. I. A full-sky Jones formalism}}.
\newblock \bibinfo{journal}{A\&A} \bibinfo{volume}{527}, \bibinfo{pages}{A106}.
\newblock \DOIprefix\doi{10.1051/0004-6361/201016082}.
\bibitem[{Smolcic et~al.(2017)Smolcic, Novak, Bondi, Ciliegi, Mooley,
  Schinnerer, Zamorani, Navarrete, Bourke, Karim, Vardoulaki, Leslie, Delhaize,
  Carilli, Myers, Baran, Delvecchio, Miettinen, Banfield, Balokovic, Bertoldi,
  Capak, Frail, Hallinan, Hao, {Herrera Ruiz}, Horesh, Ilbert, Intema, Jelic,
  Klockner, Krpan, Kulkarni, McCracken, Laigle, Middleberg, Murphy, Sargent,
  Scoville and Sheth}]{COSMOS2017}
\bibinfo{author}{Smolcic, V.}, \bibinfo{author}{Novak, M.},
  \bibinfo{author}{Bondi, M.}, \bibinfo{author}{Ciliegi, P.},
  \bibinfo{author}{Mooley, K.P.}, \bibinfo{author}{Schinnerer, E.},
  \bibinfo{author}{Zamorani, G.}, \bibinfo{author}{Navarrete, F.},
  \bibinfo{author}{Bourke, S.}, \bibinfo{author}{Karim, A.},
  \bibinfo{author}{Vardoulaki, E.}, \bibinfo{author}{Leslie, S.},
  \bibinfo{author}{Delhaize, J.}, \bibinfo{author}{Carilli, C.L.},
  \bibinfo{author}{Myers, S.T.}, \bibinfo{author}{Baran, N.},
  \bibinfo{author}{Delvecchio, I.}, \bibinfo{author}{Miettinen, O.},
  \bibinfo{author}{Banfield, J.}, \bibinfo{author}{Balokovic, M.},
  \bibinfo{author}{Bertoldi, F.}, \bibinfo{author}{Capak, P.},
  \bibinfo{author}{Frail, D.A.}, \bibinfo{author}{Hallinan, G.},
  \bibinfo{author}{Hao, H.}, \bibinfo{author}{{Herrera Ruiz}, N.},
  \bibinfo{author}{Horesh, A.}, \bibinfo{author}{Ilbert, O.},
  \bibinfo{author}{Intema, H.}, \bibinfo{author}{Jelic, V.},
  \bibinfo{author}{Klockner, H.R.}, \bibinfo{author}{Krpan, J.},
  \bibinfo{author}{Kulkarni, S.R.}, \bibinfo{author}{McCracken, H.},
  \bibinfo{author}{Laigle, C.}, \bibinfo{author}{Middleberg, E.},
  \bibinfo{author}{Murphy, E.J.}, \bibinfo{author}{Sargent, M.},
  \bibinfo{author}{Scoville, N.Z.}, \bibinfo{author}{Sheth, K.},
  \bibinfo{year}{2017}.
\newblock \bibinfo{title}{{The VLA-COSMOS 3 GHz large project: continuum data
  and source catalog release}}.
\newblock \bibinfo{journal}{A\&A} \bibinfo{volume}{602}, \bibinfo{pages}{A1}.
\newblock \DOIprefix\doi{10.1051/0004-6361/201628704}.
\bibitem[{{Taffoni} et~al.(2020){Taffoni}, {Becciani}, {Garilli}, {Maggio},
  {Pasian}, {Umana}, {Smareglia} and {Vitello}}]{CHIPP}
\bibinfo{author}{{Taffoni}, G.}, \bibinfo{author}{{Becciani}, U.},
  \bibinfo{author}{{Garilli}, B.}, \bibinfo{author}{{Maggio}, G.},
  \bibinfo{author}{{Pasian}, F.}, \bibinfo{author}{{Umana}, G.},
  \bibinfo{author}{{Smareglia}, R.}, \bibinfo{author}{{Vitello}, F.},
  \bibinfo{year}{2020}.
\newblock \bibinfo{title}{{CHIPP: INAF Pilot Project for HTC, HPC and HPDA}},
  in: \bibinfo{editor}{{Pizzo}, R.}, \bibinfo{editor}{{Deul}, E.R.},
  \bibinfo{editor}{{Mol}, J.D.}, \bibinfo{editor}{{de Plaa}, J.},
  \bibinfo{editor}{{Verkouter}, H.} (Eds.), \bibinfo{booktitle}{Astronomical
  Data Analysis Software and Systems XXIX}, p. \bibinfo{pages}{307}.
\newblock \DOIprefix\doi{10.48550/arXiv.2002.01283}.
\bibitem[{{Thompson} et~al.(1986){Thompson}, {Moran} and {Swenson}}]{Thompson}
\bibinfo{author}{{Thompson}, A.}, \bibinfo{author}{{Moran}, J.},
  \bibinfo{author}{{Swenson}, G.}, \bibinfo{year}{1986}.
\newblock \bibinfo{title}{Interferometry and Synthesis in Radio Astronomy}.
\newblock \bibinfo{publisher}{John Wiley \& Sons, New York}.
\bibitem[{Zuntz et~al.(2013)Zuntz, Kacprzak, Voigt, Hirsch, Rowe and
  S}]{im3shape}
\bibinfo{author}{Zuntz, J.}, \bibinfo{author}{Kacprzak, T.},
  \bibinfo{author}{Voigt, L.}, \bibinfo{author}{Hirsch, M.},
  \bibinfo{author}{Rowe, B.}, \bibinfo{author}{S, B.}, \bibinfo{year}{2013}.
\newblock \bibinfo{title}{{IM3SHAPE: a maximum likelihood galaxy shear
  measurement code for cosmic gravitational lensing}}.
\newblock \bibinfo{journal}{MNRAS} \bibinfo{volume}{434},
  \bibinfo{pages}{1604--1618}.
\newblock \DOIprefix\doi{10.1093/mnras/stt1125}.
\bibitem[{{Zuntz} et~al.(2018){Zuntz}, {Sheldon}, {Samuroff}, {Troxel},
  {Jarvis}, {MacCrann}, {Gruen}, {Prat}, {S{\'a}nchez}, {Choi}, {Bridle},
  {Bernstein}, {Dodelson}, {Drlica-Wagner}, {Fang}, {Gruendl}, {Hoyle}, {Huff},
  {Jain}, {Kirk}, {Kacprzak}, {Krawiec}, {Plazas}, {Rollins}, {Rykoff},
  {Sevilla-Noarbe}, {Soergel}, {Varga}, {Abbott}, {Abdalla}, {Allam}, {Annis},
  {Bechtol}, {Benoit-L{\'e}vy}, {Bertin}, {Buckley-Geer}, {Burke}, {Carnero
  Rosell}, {Carrasco Kind}, {Carretero}, {Castander}, {Crocce}, {Cunha},
  {D'Andrea}, {da Costa}, {Davis}, {Desai}, {Diehl}, {Dietrich}, {Doel},
  {Eifler}, {Estrada}, {Evrard}, {Fausti Neto}, {Fernandez}, {Flaugher},
  {Fosalba}, {Frieman}, {Garc{\'{\i}}a-Bellido}, {Gaztanaga}, {Gerdes},
  {Giannantonio}, {Gschwend}, {Gutierrez}, {Hartley}, {Honscheid}, {James},
  {Jeltema}, {Johnson}, {Johnson}, {Kuehn}, {Kuhlmann}, {Kuropatkin}, {Lahav},
  {Li}, {Lima}, {Maia}, {March}, {Martini}, {Melchior}, {Menanteau}, {Miller},
  {Miquel}, {Mohr}, {Neilsen}, {Nichol}, {Ogando}, {Roe}, {Romer}, {Roodman},
  {Sanchez}, {Scarpine}, {Schindler}, {Schubnell}, {Smith}, {Smith},
  {Soares-Santos}, {Sobreira}, {Suchyta}, {Swanson}, {Tarle}, {Thomas},
  {Tucker}, {Vikram}, {Walker}, {Wechsler} and {Zhang}}]{Zuntz18}
\bibinfo{author}{{Zuntz}, J.}, \bibinfo{author}{{Sheldon}, E.},
  \bibinfo{author}{{Samuroff}, S.}, \bibinfo{author}{{Troxel}, M.A.},
  \bibinfo{author}{{Jarvis}, M.}, \bibinfo{author}{{MacCrann}, N.},
  \bibinfo{author}{{Gruen}, D.}, \bibinfo{author}{{Prat}, J.},
  \bibinfo{author}{{S{\'a}nchez}, C.}, \bibinfo{author}{{Choi}, A.},
  \bibinfo{author}{{Bridle}, S.L.}, \bibinfo{author}{{Bernstein}, G.M.},
  \bibinfo{author}{{Dodelson}, S.}, \bibinfo{author}{{Drlica-Wagner}, A.},
  \bibinfo{author}{{Fang}, Y.}, \bibinfo{author}{{Gruendl}, R.A.},
  \bibinfo{author}{{Hoyle}, B.}, \bibinfo{author}{{Huff}, E.M.},
  \bibinfo{author}{{Jain}, B.}, \bibinfo{author}{{Kirk}, D.},
  \bibinfo{author}{{Kacprzak}, T.}, \bibinfo{author}{{Krawiec}, C.},
  \bibinfo{author}{{Plazas}, A.A.}, \bibinfo{author}{{Rollins}, R.P.},
  \bibinfo{author}{{Rykoff}, E.S.}, \bibinfo{author}{{Sevilla-Noarbe}, I.},
  \bibinfo{author}{{Soergel}, B.}, \bibinfo{author}{{Varga}, T.N.},
  \bibinfo{author}{{Abbott}, T.M.C.}, \bibinfo{author}{{Abdalla}, F.B.},
  \bibinfo{author}{{Allam}, S.}, \bibinfo{author}{{Annis}, J.},
  \bibinfo{author}{{Bechtol}, K.}, \bibinfo{author}{{Benoit-L{\'e}vy}, A.},
  \bibinfo{author}{{Bertin}, E.}, \bibinfo{author}{{Buckley-Geer}, E.},
  \bibinfo{author}{{Burke}, D.L.}, \bibinfo{author}{{Carnero Rosell}, A.},
  \bibinfo{author}{{Carrasco Kind}, M.}, \bibinfo{author}{{Carretero}, J.},
  \bibinfo{author}{{Castander}, F.J.}, \bibinfo{author}{{Crocce}, M.},
  \bibinfo{author}{{Cunha}, C.E.}, \bibinfo{author}{{D'Andrea}, C.B.},
  \bibinfo{author}{{da Costa}, L.N.}, \bibinfo{author}{{Davis}, C.},
  \bibinfo{author}{{Desai}, S.}, \bibinfo{author}{{Diehl}, H.T.},
  \bibinfo{author}{{Dietrich}, J.P.}, \bibinfo{author}{{Doel}, P.},
  \bibinfo{author}{{Eifler}, T.F.}, \bibinfo{author}{{Estrada}, J.},
  \bibinfo{author}{{Evrard}, A.E.}, \bibinfo{author}{{Fausti Neto}, A.},
  \bibinfo{author}{{Fernandez}, E.}, \bibinfo{author}{{Flaugher}, B.},
  \bibinfo{author}{{Fosalba}, P.}, \bibinfo{author}{{Frieman}, J.},
  \bibinfo{author}{{Garc{\'{\i}}a-Bellido}, J.}, \bibinfo{author}{{Gaztanaga},
  E.}, \bibinfo{author}{{Gerdes}, D.W.}, \bibinfo{author}{{Giannantonio}, T.},
  \bibinfo{author}{{Gschwend}, J.}, \bibinfo{author}{{Gutierrez}, G.},
  \bibinfo{author}{{Hartley}, W.G.}, \bibinfo{author}{{Honscheid}, K.},
  \bibinfo{author}{{James}, D.J.}, \bibinfo{author}{{Jeltema}, T.},
  \bibinfo{author}{{Johnson}, M.W.G.}, \bibinfo{author}{{Johnson}, M.D.},
  \bibinfo{author}{{Kuehn}, K.}, \bibinfo{author}{{Kuhlmann}, S.},
  \bibinfo{author}{{Kuropatkin}, N.}, \bibinfo{author}{{Lahav}, O.},
  \bibinfo{author}{{Li}, T.S.}, \bibinfo{author}{{Lima}, M.},
  \bibinfo{author}{{Maia}, M.A.G.}, \bibinfo{author}{{March}, M.},
  \bibinfo{author}{{Martini}, P.}, \bibinfo{author}{{Melchior}, P.},
  \bibinfo{author}{{Menanteau}, F.}, \bibinfo{author}{{Miller}, C.J.},
  \bibinfo{author}{{Miquel}, R.}, \bibinfo{author}{{Mohr}, J.J.},
  \bibinfo{author}{{Neilsen}, E.}, \bibinfo{author}{{Nichol}, R.C.},
  \bibinfo{author}{{Ogando}, R.L.C.}, \bibinfo{author}{{Roe}, N.},
  \bibinfo{author}{{Romer}, A.K.}, \bibinfo{author}{{Roodman}, A.},
  \bibinfo{author}{{Sanchez}, E.}, \bibinfo{author}{{Scarpine}, V.},
  \bibinfo{author}{{Schindler}, R.}, \bibinfo{author}{{Schubnell}, M.},
  \bibinfo{author}{{Smith}, M.}, \bibinfo{author}{{Smith}, R.C.},
  \bibinfo{author}{{Soares-Santos}, M.}, \bibinfo{author}{{Sobreira}, F.},
  \bibinfo{author}{{Suchyta}, E.}, \bibinfo{author}{{Swanson}, M.E.C.},
  \bibinfo{author}{{Tarle}, G.}, \bibinfo{author}{{Thomas}, D.},
  \bibinfo{author}{{Tucker}, D.L.}, \bibinfo{author}{{Vikram}, V.},
  \bibinfo{author}{{Walker}, A.R.}, \bibinfo{author}{{Wechsler}, R.H.},
  \bibinfo{author}{{Zhang}, Y.}, \bibinfo{year}{2018}.
\newblock \bibinfo{title}{{Dark Energy Survey Year 1 Results: Weak Lensing
  Shape Catalogues}}.
\newblock \bibinfo{journal}{MNRAS} \bibinfo{volume}{481},
  \bibinfo{pages}{1149--1182}.
\newblock \DOIprefix\doi{10.1093/mnras/sty2219}.

\end{thebibliography}

\end{document}